\documentclass{article}
\usepackage{graphicx} 
\usepackage[numbers]{natbib}
\usepackage{array}
\usepackage{booktabs}
\usepackage{threeparttable}
\usepackage{multirow}
\usepackage{url}
\usepackage{lipsum}
\usepackage{amssymb}
\usepackage{amsmath}
\usepackage{mathrsfs}
\usepackage[hidelinks]{hyperref}
\usepackage{siunitx}
\usepackage{float}
\usepackage{subcaption}
\usepackage{pdflscape}
\usepackage{rotating}
\usepackage{multifootnote}
\usepackage{authblk}

\usepackage{geometry}
\newcommand{\virgolette}[1]{``#1''}

\title{\textbf{Advances in Ontology--Based Mining of Adverse Drug Reactions} \\[1ex] \large \textbf{\textit{Preprint}}}
\author[1]{Kenenisa Tadesse Dame\footnote{Correspondence to: \url{kenenisatadesse.dame@phd.unipd.it}}}
\author[1]{Pietro Belloni}
\author[2]{Ugo Moretti}
\author[2]{Fabio Scapini}
\author[2]{Marco Tuccori}
\author[1]{Alessandra R. Brazzale}
\affil[1]{Department of Statistical Sciences, University of Padova, Italy}
\affil[2]{Department of Diagnostics and Public Health, University of Verona, Italy}

\date{Version: October 2026}

\begin{document}

\maketitle

\section*{Abstract}
Post--marketing pharmacovigilance is essential for identifying adverse drug reactions (ADRs) that elude  detection during pre--marketing clinical trials.  This study explores a novel approach that integrates an adverse event (AE) ontology into a zero--inflated negative binomial model to improve ADR detection. By accounting for the biological similarities among correlated AEs and addressing the excess of zero counts, this method more effectively disentangles AE associations. Statistical significance is evaluated using a permutation--based maximum statistic that preserves AE correlations within individual reports. Simulations and an application to real data from the Veneto drug safety database demonstrate that the ontology--based model consistently outperforms classical models such as the Gamma--Poisson Shrinker (GPS). For post--selection inference, we furthermore explore a data thinning technique for convolution--closed families, enabling the creation of independent training and validation datasets while retaining all drug--AE pairs. This approach is compared with conventional random train/test splitting, which may leave some drugs or AEs absent from one subset, and stratified splitting, which requires expanding aggregated counts into individual instances. The data--thinning technique and stratified splitting yield very similar results, with stratified splitting showing a slight benefit, and both clearly outperform random splitting in ensuring reliable and consistent model evaluation.

\vspace{0.5cm}

\noindent
\textbf{Keywords}: Data Thinning, Empirical Bayes, Pharmacovigilance, Post--Selection Inference, Zero--Inflated Model

\section{Motivation and background}\label{sec1}
An adverse event (AE) is any untoward medical occurrence in a patient receiving a medicinal product, regardless of causality. Adverse drug reactions (ADRs) are a subset of AEs that are harmful, unintended and reasonably attributed to the drug. Although these terms are sometimes used interchangeably in the literature, all ADRs are AEs, but not vice versa.

Adverse drug reactions (ADRs) remain a leading cause of global morbidity and mortality despite considerable progress in pharmaceutical development and drug approvals~\cite{who2004,european2009benefit}. This persistent burden arises because pre--marketing clinical trials, though essential for evaluating drug safety, are inherently constrained by limited sample sizes, restricted participant diversity and relatively short study durations. As a result, rare, long--term or population--specific ADRs often go undetected during preclinical and clinical evaluations~\cite{efpia2023development,who2023clinicaltrials,fda2018postmarket}. Post--marketing pharmacovigilance is therefore crucial  for detecting and monitoring ADRs that escape detection during the early studies. This process largely depends on spontaneous reporting systems (SRSs), in which Individual Case Safety Reports (ICSRs) are voluntarily submitted by healthcare professionals, patients or marketing authorization holders. ICSRs are compiled in national databases, such as the Italian \textit{Rete Nazionale di Farmacovigilanza}, and shared with regional and global repositories, including EudraVigilance and the WHO’s VigiBase™, enabling large--scale signal detection and comprehensive drug safety monitoring.

SRSs are particularly effective for detecting rare or unexpected ADRs, especially during the early post--marketing period~\cite{lu2009information,zorych2013disproportionality}. However, they suffer from underreporting, reporting biases, missing data, duplicate entries and lack of denominator information~\cite{hazell2006,wallenstein2001temporal,hartnell2004replication,han2025,sign}. Despite these challenges, SRSs remain essential, as no alternative data source fully captures the breadth of ADR occurrences~\cite{Stephenson2007}. To date, numerous statistical methods have been applied to detect ADR signals. These include, among others, measures of effects such as the reporting odds ratio (ROR) and the proportional reporting ratio (PRR), but also more elaborated approaches such as Bayesian confidence propagation neural networks (BCPNN) and Gamma--Poisson shrinkage (GPS)~\cite{pham2019comparison}.  All these methods compare the observed frequency of drug--event combinations with their expected frequencies in spontaneous reporting systems~\cite{evans2001use,van2002comparison,bate1998bayesian,orre2000bayesian,noren2006extending,Dumouchel1999, belloni2024age}.  In addition, they all assume independence among AEs.

However, AEs are  de facto intrinsically linked. Adverse events are classified according to the Medical Dictionary for Regulatory Activities (MedDRA, Version 27.1), which uses a multi--level hierarchical structure. Furthermore, many AEs go unreported for certain drugs, challenging the analysis of pharmacovigilance data because of the high prevalence of excess zero counts.  Explicitly integrating the ontology of the AEs into the estimation process would allow sharing the biological similarities among related AEs, which in turn enhances the accuracy of true AE signal detection while minimizing false positives.  This challenge is addressed by the zero--inflated negative binomial (ZINB) model~\cite{lee2023evaluation,yau2003zero}, which accounts for zero counts while using shared information among AEs within the same group.

Furthermore, traditional sample splitting at the drug--AE pair level poses challenges for validating pharmacovigilance models, as it may exclude certain drugs or adverse events from either the training or validation sets, thereby undermining performance assessment. To address this, we introduce a data thinning based on convolution--closed distributions, which partitions counts while preserving their underlying structure~\cite{rasines2022splitting,JMLR:v25:23-0446}. This enables robust model validation that respects the inherent relationships in SRS data. The remainder of this article is structured as follows.  Section~\ref{sec:stat_modeling} introduces notations and reviews Gamma--Poisson shrinkage. Section~\ref{sec:ontology_model} presents the ontology--based signal detection model. Section~\ref{sec:model_validation} details an innovative model validation setting using post--selection inference. Sections ~\ref{sec:numerical_results}, \ref{sec:model_performance} and~\ref{sec:applications_discussion} report numerical assessment, real data applications, and discussion.

\section{Modeling signal detection in pharmacovigilance}
\label{sec:stat_modeling}

\subsection{Notation and definition}
\label{subsec:defn}
Suppose that, over a specified time period and for the same adverse events (AEs), we observe $J>0$ events for $I>0$ drugs.  The data can be summarized through an $I \times J$ contingency table (see Table~\ref{tab:cont}), \begin{table}[t!]
\centering
\renewcommand{\arraystretch}{1.15}
\caption{$I \times J$ contingency table of drug--adverse event (AE) pairs.}
\begin{tabular}{cccccc}
\toprule
\textbf{Drug \(\backslash\) AE} 
& \textbf{AE\(_1\)} 
& \textbf{AE\(_2\)} 
& \(\cdots\) 
& \textbf{AE\(_J\)} 
& \textbf{Total} \\ 
\midrule

\(\textbf{X}_1\) 
& \(n_{11}\) 
& \(n_{12}\) 
& \(\cdots\) 
& \(n_{1J}\) 
& \(n_{1.} = \sum_{j=1}^{J} n_{1j}\) \\

\(\textbf{X}_2\) 
& \(n_{21}\) 
& \(n_{22}\) 
& \(\cdots\) 
& \(n_{2J}\) 
& \(n_{2.} = \sum_{j=1}^{J} n_{2j}\) \\

\(\vdots\) 
& \(\vdots\) 
& \(\vdots\) 
& \(\vdots\) 
& \(\vdots\) 
& \(\vdots\) \\

\(\textbf{X}_I\) 
& \(n_{I1}\) 
& \(n_{I2}\) 
& \(\cdots\) 
& \(n_{IJ}\) 
& \(n_{I.} = \sum_{j=1}^{J} n_{Ij}\) \\

\midrule
\textbf{Total} 
& \(n_{.1} = \sum_{i=1}^{I} n_{i1}\) 
& \(n_{.2} = \sum_{i=1}^{I} n_{i2}\) 
& \(\cdots\) 
& \(n_{.J} = \sum_{i=1}^{I} n_{iJ}\) 
& \(n_{..} = \sum_{i=1}^{I} \sum_{j=1}^{J} n_{ij}\) \\

\bottomrule
\end{tabular}
\label{tab:cont}
\end{table}
 where $n_{ij}$ counts the number of reports which mention both drug $i$ and AE $j$.  The marginal totals are given by

\begin{equation}
n_{i.} = \sum_{j=1}^{J} n_{ij}, \quad 
n_{.j} = \sum_{i=1}^{I} n_{ij}, \quad 
n_{..} = \sum_{i=1}^{I} \sum_{j=1}^{J} n_{ij},
\nonumber
\end{equation}
with $n_{i.}$ being the total count for drug $i$, $n_{.j}$ the total count for AE $j$, and $n_{..}$ the overall total.  We can readily define the expected frequency $E_{ij}$ for the $ij$-th drug--AE pair assuming no association between drugs and AEs, and the corresponding relative reporting rate, or reporting ratio (RR for short), $\lambda_{ij}$ as
\begin{equation}
E_{ij} = \frac{n_{i.} n_{.j}}{n_{..}} \quad\quad \text{and} \quad\quad \lambda_{ij} =RR_{ij}= \frac{n_{ij}}{E_{ij}}.
\label{eq:Eij}
\end{equation}
The relative reporting ratio is intuitive: for example, $\text{RR}_{ij} = 1{,}000$ means that AE $j$ occurred 1,000 times more frequently with drug $i$ than expected under independence. However, this measure suffers from high sampling variability when counts are small. For instance, both $n_{ij}=1, E_{ij}=0.001$ and $n_{ij}=100, E_{ij}=0.1$ yield $\text{RR}_{ij}=1{,}000$, yet their statistical reliability differs substantially. To address this limitation, an equivalent measure has been proposed~\cite{church1990word}: the mutual information statistic, defined as
\[
I_{ij} = \log_2(\text{RR}_{ij}),
\]
which quantifies the strength of association on the binary log scale.
A large observed value of $RR_{ij}$ (or equivalently $I_{ij}$) suggests a potential drug safety issue.

\subsection{Gamma--Poisson shrinkage model}
The Gamma--Poisson shrinkage (GPS) model is a Bayesian hierarchical framework widely employed for signal detection in spontaneous reporting systems of pharmacovigilance~\cite{Dumouchel1999}.
It models the observed count \( n_{ij} \) which represents the number of reports of drug \( i \) associated with adverse event \( j \), as an independent Poisson random variable

\begin{equation}
N_{ij} \sim \text{Poisson}(\mu_{ij}),\quad\text{with}\quad\mu_{ij} = \lambda_{ij}E_{ij}, 
\label{eq:lam}
\end{equation}
where \( E_{ij} \) and \( \lambda_{ij} \) are the expected count and the relative reporting rate, respectively, as explained in \eqref{eq:Eij}.
 Instead of treating each \( \lambda_{ij} \) as an independent parameter, the GPS model introduces shrinkage by assuming \( \lambda_{ij} \) follows a two--component mixture of gamma distributions with parameter \(\theta= (\alpha_1, \beta_1, \alpha_2, \beta_2, \omega) \),
\begin{equation}
    \lambda_{ij} \sim\omega g(\lambda_{ij}; \alpha_1, \beta_1) + (1 - \omega) g(\lambda_{ij}; \alpha_2, \beta_2),
\label{eq:prior_gps}
\end{equation} where \( \omega \in (0,1) \) is the probability that \( \lambda_{ij} \) comes from the first component of the mixture and \( (\alpha_k, \beta_k) > 0 \) for \( k = 1, 2 \), represent the shape and rate parameters of the two gamma components, while \(g(\lambda; \alpha, \beta) = \frac{\beta^{\alpha}}{\Gamma(\alpha)} \lambda^{\alpha - 1} e^{-\beta \lambda}
\)  denotes the gamma probability density function with mean \( \alpha / \beta \) and variance \( \alpha / \beta^2 \). By integrating over $\lambda_{ij}$ in model~\eqref{eq:lam} using \eqref{eq:prior_gps}, the marginal distribution of the count $n_{ij}$,
\begin{equation}
\text{Pr}(N_{ij} = n_{ij} \mid \theta, E_{ij}) =
\omega  \text{NB}\left(n_{ij};\alpha_1, \frac{E_{ij}}{E_{ij} + \beta_1}\right) + (1 - \omega)  \text{NB}\left(n_{ij};\alpha_2, \frac{E_{ij}}{E_{ij} + \beta_2}\right),
\label{ZINB-model_alt}
\end{equation}
turns out to be a mixture of two negative binomial models, where \(\text{NB}(n_{ij};\alpha,E_{ij}/(E_{ij}+\beta))\) is the negative binomial probability evaluated at $n_{ij}$ with shape parameter $\alpha$ and parameter $E_{ij}/(E_{ij}+\beta)$.

The hyperparameters \(\theta\) are estimated empirically by maximizing the marginal likelihood over all drug--event pairs\((i,j)\)
\[
\hat{\theta} = \arg \max_{\theta} L(\theta \mid n) = \arg \max_{\theta} \prod_{i,j} \Pr(N_{ij}=n_{ij} \mid \theta, E_{ij})
\]
where \(n = \{n_{ij} : i = 1, \ldots, I; j = 1, \ldots, J\}\) denotes the vector of observed counts  
and \(L(\theta \mid n)\) is the corresponding likelihood.  This maximization is performed iteratively in the five--dimensional parameter space, usually starting from the initial values

\[
\alpha_1 = 0.2, \quad \beta_1 = 0.1, \quad \alpha_2 = 2, \quad \beta_2 = 4, \quad \omega = \tfrac{1}{3}
\] until convergence~\cite{Dumouchel1999}, yielding the maximum likelihood estimate (MLE) \(\hat{\theta} = (\hat{\alpha}_1, \hat{\beta}_1, \hat{\alpha}_2, \hat{\beta}_2, \hat{\omega})\). The relative reporting rates \(\lambda_{ij}\) for each AE are then estimated using an empirical Bayes approach. 
The empirical prior density for \(\lambda_{ij}\), 

\[\pi(\lambda_{ij} \mid \hat{\theta}) = \hat{\omega} g(\lambda_{ij} \mid \hat{\alpha}_1, \hat{\beta}_1) + (1 - \hat{\omega}) g(\lambda_{ij} \mid \hat{\alpha}_2, \hat{\beta}_2),\]
with parameters replaced by their MLEs, is combined with the Poisson likelihood 
\[
L(\lambda_{ij} \mid n_{ij} ) = \frac{\exp(-E_{ij} \lambda_{ij})(E_{ij} \lambda_{ij})^{n_{ij}}}{n_{ij}!}
\]
for $\lambda_{ij}$ given $n_{ij}$. The  posterior density of \(\lambda_{ij}\) is
\begin{equation*}
    \pi(\lambda_{ij} \mid n_{ij}) = 
    \omega^{*}g\left(\lambda_{ij}; \hat{\alpha}_{1} + n_{ij}, \, \hat{\beta}_{1} + E_{ij}\right)  
    + \left(1 - \omega^{*}\right) g\left(\lambda_{ij}; \hat{\alpha}_{2} + n_{ij}, \, \hat{\beta}_{2} + E_{ij}\right) ,
    \label{eq:post_gps}
\end{equation*}
and it is also a mixture of gamma densities,
where \(\omega^*\) is the posterior probability of \(\lambda_{ij}\) coming from the first component of the mixture
\[
\omega^* = 
\frac{
\hat{\omega} \cdot \text{NB}\!\left(n_{ij}; \hat{\alpha}_1, \frac{E_{ij}}{E_{ij} + \hat{\beta}_1}\right)
}{
\hat{\omega} \cdot \text{NB}\!\left(n_{ij}; \hat{\alpha}_1, \frac{E_{ij}}{E_{ij} + \hat{\beta}_1}\right)
+ (1 - \hat{\omega}) \cdot \text{NB}\!\left(n_{ij}; \hat{\alpha}_2, \frac{E_{ij}}{E_{ij} + \hat{\beta}_2}\right)
}.\]
The posterior means of \(\lambda_{ij}\) and \(\log \lambda_{ij}\) are 

\[
\mathrm{E}\left(\lambda_{ij} \mid n_{ij}\right) = 
\omega^{*} \cdot \frac{\hat{\alpha}_{1} + n_{ij}}{\hat{\beta}_{1} + E_{ij}} 
+ (1 - \omega^{*}) \cdot \frac{\hat{\alpha}_{2} + n_{ij}}{\hat{\beta}_{2} + E_{ij}},
\]
and

\[
\mathrm{E}(\log \lambda_{ij} \mid n_{ij}) = 
\omega^{*} \left[ \Psi(\hat{\alpha}_{1} + n_{ij}) - \log(\hat{\beta}_{1} + E_{ij}) \right] 
+ (1 - \omega^{*}) \left[ \Psi(\hat{\alpha}_{2} + n_{ij}) - \log(\hat{\beta}_{2} + E_{ij}) \right],
\]where \(\Psi(\cdot)\) is the digamma function, the derivative of \(\log \Gamma(\cdot)\).

The quantity
\[
\mathrm{EB}\log_2{\lambda}_{ij} = \mathbb{E}[\log_2(\lambda_{ij}) \mid n_{ij}] = \frac{\mathbb{E}[\log(\lambda_{ij}) \mid n_{ij}]}{\log 2}
\]
is the empirical Bayesian (EB) counterpart of \(\log_2 (\mathrm{RR}_{ij})\). To express this estimate on the same scale as the relative reporting rate, we apply the exponential transformation

\[
\mathrm{EBGM}_{ij} = 2^{\mathrm{EB}\log_2{\lambda}_{ij}}
\]
where EBGM stands for \emph{empirical Bayes geometric mean}.

The posterior distribution of \(\lambda_{ij}\) allows calculation of credibility intervals for the EBGM measure using its percentiles. For instance, a 99\% credibility interval is defined by the 1st and 99th percentiles, \(\lambda_{ij}^{0.01}\) and \(\lambda_{ij}^{0.99}\). If the lower bound exceeds 1, this indicates strong evidence of a potential drug--adverse event association. These percentiles are efficiently estimated using the \texttt{quantBisect()} function from the R package \texttt{openEBGM}, which implements a bisection algorithm to accurately estimate the posterior quantiles~\cite{canida2017openebgm}.

\section{Ontology--based signal detection}
\label{sec:ontology_model}
\subsection{Model definition}
Building on the contingency table framework for individual adverse events (Table~\ref{tab:cont}), we extend the analysis to groups of related AEs defined by the MedDRA ontology, which hierarchically organizes adverse events based on biological and clinical relationships. Each AE group is represented by a separate contingency table, allowing aggregation of related events.

The \textit{zero--inflated Gamma--Poisson shrinker with AE ontology} (zGPS.AO, for short) method~\cite{zhao2023mining} is a hierarchical empirical Bayesian framework  designed for analyzing AE data in drug safety surveillance, which addresses the two key challenges of excess zeros and overdispersion in count data. 
In zGPS.AO, each AE group is modeled independently. For each AE group, the count \( n_{ij} \) follows a Poisson distribution
\begin{equation}
N_{ij} \sim \text{Poisson}(E_{ij} \lambda_{ij}),
\label{eq:n:poi}
\end{equation}
where $E_{ij}$ is as in \eqref{eq:Eij} while the AE--level relative reporting rate 
\begin{equation}
\lambda_{ij} \sim \begin{cases}
\delta_0, & \text{with probability } p_i \\
g(r, \mu_i / r), & \text{with probability } 1- p_i
\end{cases}
\label{eq:lambda}
\end{equation}
is modeled as a mixture of a point mass~\cite{khuri2004applications} $\delta_0$ at 0  and a gamma distribution with shape parameter $r$ and scale parameter $\mu_i/r$ with mixing probabilities $p_i$ and $1-p_i$, respectively.  Here, $p_i$ represents the proportion of excess zeros for drug $i$, while the mean $\mu_i$ of the gamma distribution defines the reporting rate of drug $i$.  All drugs share the same shape parameter \( r \), but may have different values for \( \mu_i \) and $p_i$.
By integrating over $\lambda_{ij}$ in model~\eqref{eq:n:poi} using \eqref{eq:lambda}, the marginal distribution of $n_{ij}$ is
\begin{equation}
\text{Pr}(N_{ij} = n_{ij} \mid r, p_i, E_{ij} \mu_i) =
\begin{cases}
p_i + (1 - p_i) \text{NB}(0 \mid r, E_{ij} \mu_i), & \text{if } n_{ij} = 0 \\
(1 - p_i) \text{NB}(n_{ij} \mid r, E_{ij} \mu_i), & \text{if } n_{ij} > 0,
\end{cases}
\label{ZINB-model}
\end{equation}
and turns out to be a zero-inflated negative binomial (ZINB) model, where $\text{NB}(n_{ij} \mid r, E_{ij} \mu_i)$ is the negative binomial probability evaluated at $n_{ij}$ with dispersion parameter \( r \) and mean \( E_{ij}\mu_i \).

We are interested in the AE group--level reporting rate 
\begin{equation}
s_i = E(\lambda_{ij}) = (1 - p_i) \mu_i
\label{eq:s}
\end{equation}
for drug $i$.  The higher the value of $s_i$, the more evidence we have that the AE group is associated with drug~$i$. Mathematical details are provided in Appendix~\ref{app:mathdetails}.

\subsection{Model fitting and inference}
\label{sec:methods}
\noindent
Suppose there are $K>0$ AE groups of interest. The zGPS.AO model is fitted separately to each AE group by combining maximum likelihood (ML) and empirical Bayes estimation in a two-step procedure.

We first estimate the group--level parameters $(p_1, \dots,$ $p_I,\mu_1, \dots, \mu_I, r)$ by maximizing the likelihood of model~\eqref{ZINB-model}.  The estimated group--level reporting rate \( \hat s_i = (1-\hat p_i)\hat\mu_i \) for drug \( i \) accounts for group--specific variations in AE reporting.  
The AE--level relative reporting rates, $\lambda_{ij}$, are then estimated by empirical Bayes.  The prior densities for \(\lambda_{ij}\)
\[
\pi(\lambda_{ij} \mid \hat p_i, \hat \mu_i, \hat r) = \hat p_i \delta_{\lambda_{ij}} + (1 - \hat p_i) g(\lambda_{ij} \mid \hat r, \hat \mu_i / \hat r),
\]with the AE group--level parameters replaced by their MLEs, are combined with the Poisson likelihood 
\[
L(\lambda_{ij} \mid n_{ij} ) = \frac{\exp(-E_{ij} \lambda_{ij})(E_{ij} \lambda_{ij})^{n_{ij}}}{n_{ij}!}, \quad \lambda_{ij} \geq 0,
\]
for $\lambda_{ij}$ given $n_{ij}$.

For $n_{ij} = 0$, the posterior density
\begin{equation}
\pi(\lambda_{ij} \mid n_{ij} = 0) = \hat{\pi}_{ij} \delta_{\lambda_{ij}} + (1 - \hat{\pi}_{ij}) g\left(\lambda_{ij} \mid \hat r, \frac{\hat\mu_i}{\hat r + E_{ij} \hat\mu_i}\right),
\nonumber
\end{equation}
is a mixture of a point mass at $\lambda_{ij}$ and a gamma density, where
\begin{equation}
\hat{\pi}_{ij} = \frac{\hat p_i}{\hat p_i + (1 - \hat p_i) \left(\frac{\hat r}{\hat r + E_{ij} \hat\mu_i}\right)^{\hat r}}
\nonumber
\end{equation}
represents the posterior probability that $n_{ij} = 0$.
For $n_{ij} > 0$, the posterior density
\begin{equation}
\pi(\lambda_{ij} \mid n_{ij} > 0) = g\left(\lambda_{ij} \bigm| \hat r + n_{ij}, \frac{\hat \mu_i}{\hat r + E_{ij} \hat\mu_i}\right)
\nonumber
\end{equation}
is a gamma distribution with shape parameter $\hat{r} + n_{ij}$ and scale parameter $\frac{\hat{\mu}i}{\hat{r} + E{ij}\hat{\mu}_i}$.
%
The empirical Bayes estimate of the AE--level relative reporting rate
\begin{equation}
\hat\lambda_{ij} = E(\lambda_{ij} \mid n_{ij}) = 
\begin{cases}
(1 - \hat{\pi}_{ij}) \frac{\hat\mu_i \hat r}{\hat r + E_{ij} \hat\mu_i}, & \text{if } n_{ij} = 0, \\
\frac{\mu_i (\hat r + n_{ij})}{\hat r + E_{ij} \hat\mu_i}, & \text{if } n_{ij} > 0.
\end{cases} 
\end{equation}
is the posterior mean of $\lambda_{ij}$.
For full derivations of the posterior distribution and its mean, see Appendix~\ref{app:posterior} and Appendix~\ref{app:posteriormean}, respectively.

%
\subsection{Testing for significance}
\noindent
Multiple drug--AE associations can then be tested for significance using the maximum test statistic approach which uses
$$ \text{maxS} = \max_{i,k} \hat s_{i}^k, \quad i = 1, \ldots, I, \ k = 1, \ldots, K, $$
the largest observed relative reporting rate $s_{i}^k$ taken across all $I$ drugs and all $K$ adverse event groups.  The null distribution assuming $\lambda_{ij}=1$ for all drug--AE pairs is analytically intractable and hence estimated via permutation.  
In such drug safety dataset, the permutation test must account for the association among adverse events (AEs) reported within the same individual report. To preserve this inherent interdependence structure, all AEs reported in a single report are treated as a single set and permutations are performed at the level of these AE sets.

To illustrate, consider two reports. The first report contains a single drug, \textit{$X_1$}, and three AEs: \textit{$AE_1$}, \textit{$AE_2$}, and \textit{$AE_3$}. The second report contains two drugs, \textit{$X_2$} and \textit{$X_3$}, and two AEs: \textit{$AE_4$} and \textit{$AE_5$}. In a permuted dataset, the AE set \((\textit{$AE_1$}, \textit{$AE_2$}, \textit{$AE_3$})\) may be reassigned to the drug set \((\textit{$X_1$}, \textit{$X_3$})\), while the AE set \((\textit{$AE_4$}, \textit{$AE_5$})\) may be reassigned to \textit{$X_1$}. This approach preserves the association among AEs within each report while disrupting the original drug--AE associations.

For each permuted dataset, the maximum score statistic, \(\max S\), is calculated. By generating \(N\) such permutations (e.g., $N = 4000$), an empirical null distribution comprising \(N + 1\) values, including that of the observed dataset, is obtained. This distribution facilitates the assessment of the statistical significance of the observed \(\max S\) by appropriately accounting for the dependence structure inherent in spontaneous reporting data.
The significance of a specific association between drug $i$ and AE group $k$ is tested using the $q$-value, that is, the $p$-value computed for $\hat s_i^k$ obtained from the null distribution of maxS suitably adjusted to account for multiple comparisons~\cite{huang2011likelihood,ding2020evaluation,thissen2002quick}.
A small $q$-value suggests a strong drug--AE group association.  
This approach not only controls for multiple testing but also allows for adjusting individual relative reporting rates (RRs) using the maximum \(\lambda_{ij}\) across all drug--AE pairs.

\section{Model Validation}
\label{sec:model_validation}
\subsection{Data thinning}
\begin{figure}[t]
    \centering
\includegraphics[width=0.9\linewidth]{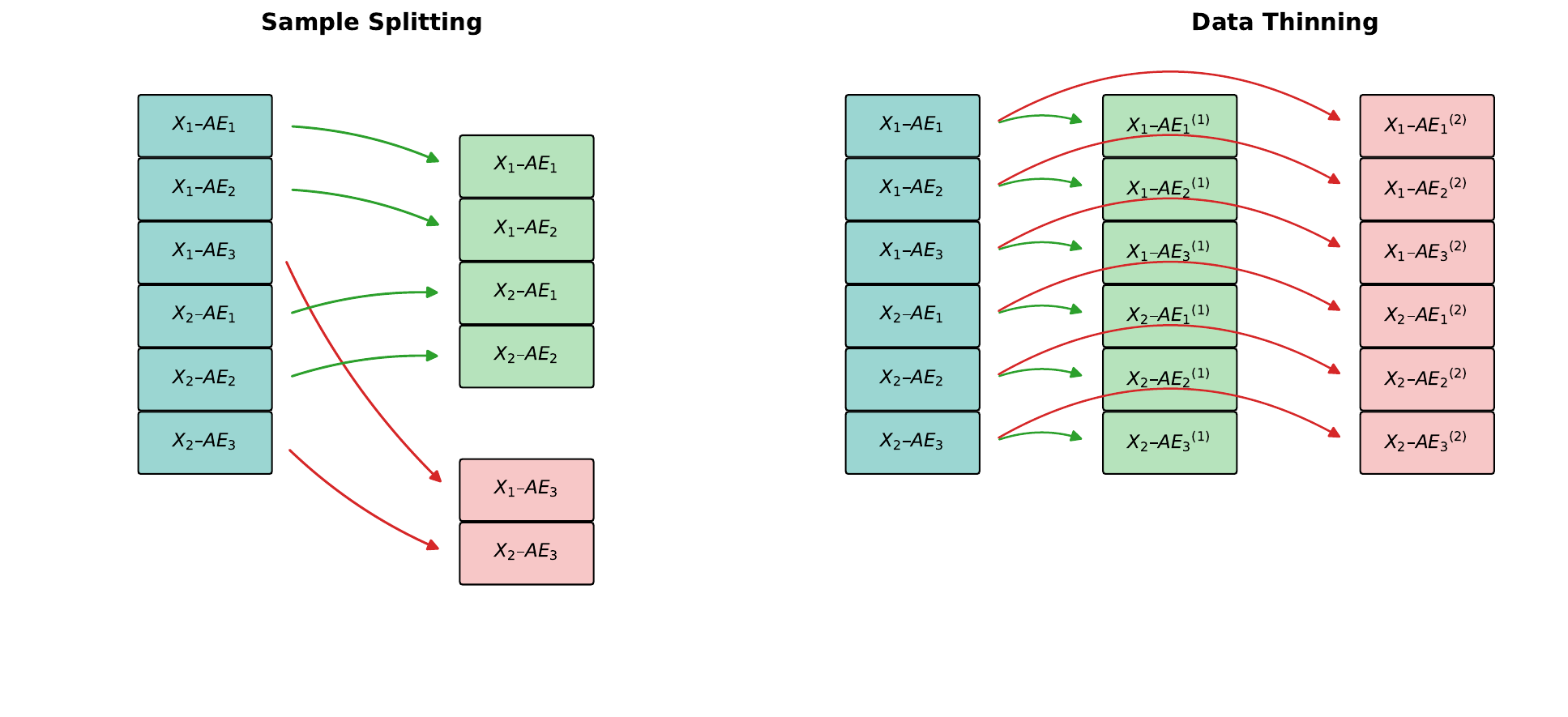}
    \caption{Left: Sample splitting randomly assigns each observation to either a training or a validation set. Right: data thinning randomly thins each observation into two parts.}
    \label{fig:data_thin}
\end{figure}

Sample splitting is a common method for model validation, where a dataset, such as counts of drug--AE pairs, is divided at random into separate training and validation subsets (see Figure~\ref{fig:data_thin}, Left Panel)~\cite{cox1975note,hastie2009elements}. This allows models to be trained on one subset and evaluated on the other. However, in pharmacovigilance, where data are aggregated counts per drug--AE pair, random splitting at the pair level often causes some drugs or AEs to appear only in one subset. This absence prevents models from learning or validating predictions for those pairs, reducing overlap and weakening inference. As a result, conventional splitting limits the robustness of model evaluation and impairs reliable signal detection.
 To address the challenges of sample splitting in pharmacovigilance, we employ convolution--closed families of probability distributions, which have the property that the sum of two independent random variables from the family also belongs to the same family, as shown Figure~\ref{fig:data_thin}, Right Panel~\cite{rasines2022splitting,JMLR:v25:23-0446,Dharamshi02012025}. Formally, a family of distributions \(\{F_\lambda : \lambda \in \Lambda\}\) is convolution-closed if, for any independent random variables \(N^{(1)} \sim F_{\lambda_1}\) and \(N^{(2)} \sim F_{\lambda_2}\), their sum
\begin{equation}
N^{(1)} + N^{(2)} \sim F_{\lambda_1 + \lambda_2},
\label{subsec:con-clos}
\end{equation}
provided that \(\lambda_1 + \lambda_2 \in \Lambda\). This family also satisfies the \textit{linear expectation property}: for \(N \sim F_\lambda\), the expectation \(\mathbb{E}[N]\) is linear in \(\lambda\). Since

\begin{equation}
\mathbb{E}[N^{(1)} + N^{(2)}] = \mathbb{E}[N^{(1)}] + \mathbb{E}[N^{(2)}],
\label{subsec:exp}
\end{equation}
and \(N^{(1)} + N^{(2)} \sim F_{\lambda_1 + \lambda_2}\), the expectation depends additively on \(\lambda\). This property enables a stable and principled approach to thinning counts in pharmacovigilance data without loss of information.

Let \( N \sim F_\lambda \) be a count variable from a convolution--closed family, with observed realization \( n_{ij} \). We aim to thin \( n_{ij} \) into two  components \( N^{(1)} \sim F_{\varepsilon\lambda} \) and \( N^{(2)} \sim F_{(1 - \varepsilon)\lambda} \), for \( \varepsilon \in (0,1) \) such that \( \varepsilon\lambda, (1-\varepsilon)\lambda \in \Lambda \).
To connect this to the observed count, we apply a method called \textit{data thinning}, where
\begin{equation}
N^{(1)} \mid N^{(1)} + N^{(2)} = n_{ij} \sim G_{\varepsilon\lambda, (1-\varepsilon)\lambda, n_{ij}},
\label{subsec:cond}
\end{equation}
and \( G_{\lambda_1, \lambda_2, n} \) is the distribution of one component given the sum of the two based on the convolution property. For example, if \( F_\lambda \) is Poisson, then \( G_{\lambda_1, \lambda_2, n} \) is binomial with parameters \( (n, \lambda_1 / (\lambda_1 + \lambda_2)) \)~\cite{jorgensen1992exponential}.

After drawing \( N^{(1)} \), the complementary component will be \( N^{(2)} = n_{ij} - N^{(1)} \) and the pair \( (N^{(1)}, N^{(2)}) \) serves as training and validation counts.
The thinned counts satisfy \( N^{(1)} \sim F_{\varepsilon\lambda} \) and \( N^{(2)} \sim F_{(1-\varepsilon)\lambda} \). If \( F_\lambda \) satisfies the linear expectation property, then their expectations scale proportionally as
\[
\mathbb{E}[N^{(1)}] = \varepsilon \mathbb{E}[N] \quad \text{and} \quad \mathbb{E}[N^{(2)}] = (1 - \varepsilon) \mathbb{E}[N].
\]

\subsection{Application to pharmacovigilance }

We assume the count of an AE group  \( n_{ij} \) modeled as in \eqref{eq:n:poi}. We define \(\varepsilon \in (0,1)\) as the proportion of observed counts allocated for training, with the remaining proportion \(1 - \varepsilon\) used for validation or testing (e.g., $\varepsilon = 0.7$ means 70\% training and 30\% validation). By convolution closure, there exist independent random variables
\begin{equation} \label{eq:poisson_split}
N_{ij}^{(1)} \sim \text{Poisson}(\varepsilon E_{ij}\lambda_{ij}), \quad
N_{ij}^{(2)} \sim \text{Poisson}((1-\varepsilon)E_{ij}\lambda_{ij}),
\end{equation}
such that
\[
N_{ij}^{(1)} + N_{ij}^{(2)} \sim \text{Poisson}(E_{ij}\lambda_{ij}), \quad N_{ij}^{(1)} \perp N_{ij}^{(2)}.
\] This approach thins each observed count into  training and validation components without data loss, preserving the original data structure.
Since the observed counts are fixed, we implement this thin conditionally as described in (\ref{subsec:cond}):

\[
N_{ij}^{(1)} \mid n_{ij} \sim \text{Binomial}(n_{ij}, \varepsilon), \quad N_{ij}^{(2)} = n_{ij} - N_{ij}^{(1)}.
\]
Moreover, since thinning is conditional on fixed observed counts, all expectations are also conditional. \[
\mathbb{E}[N_{ij}^{(1)} \mid n_{ij}] = \varepsilon n_{ij}, \quad
\mathbb{E}[N_{ij}^{(2)} \mid n_{ij}] = \mathbb{E}[n_{ij} - N_{ij}^{(1)} \mid n_{ij}] = n_{ij} - \varepsilon n_{ij} = (1 - \varepsilon) n_{ij}.
\]

\section{Application to mock and real data}
\label{sec:numerical_results} 
\subsection{Simulation I}
We began by conducting a series of small simulation studies to assess how the zGPS.AO model performs across different group sizes under varying dispersion structures. We picked one large AE group from MedDRA and three drugs that had the most reports from the Veneto drug safety dataset. We calculated the expected counts $E_{ij}$ as in \eqref{eq:Eij}, and found the true group--level parameters by fitting a zero--inflated negative binomial (ZINB) model to the data. The simulation parameters were $r = 5.72$; $p = (0.133, 0.350, 0.0669)$; $\mu = (0.933, 1.30, 0.816$).

For each group size \(J = 10, 20, 40, 80\), \(\lambda_{ij}\) were generated from Equation~\eqref{eq:lambda}, and counts \(n_{ij}\) from Equation~\eqref{eq:n:poi}, with \(\hat{\lambda}_{ij}\) and \(\hat{s}_i\) recorded over 1000 iterations to test the impact of group size on performance. This setup was defined as the \virgolette{standard} simulation. To assess robustness, a \virgolette{no--dispersion} scenario (i.e., no negative binomial dispersion) was also considered, in which counts were generated from a zero--inflated Poisson model (\(r \to \infty\)) while keeping all other parameters unchanged.

\begin{figure}[t]
    \centering
    \includegraphics[width=0.7\textwidth]{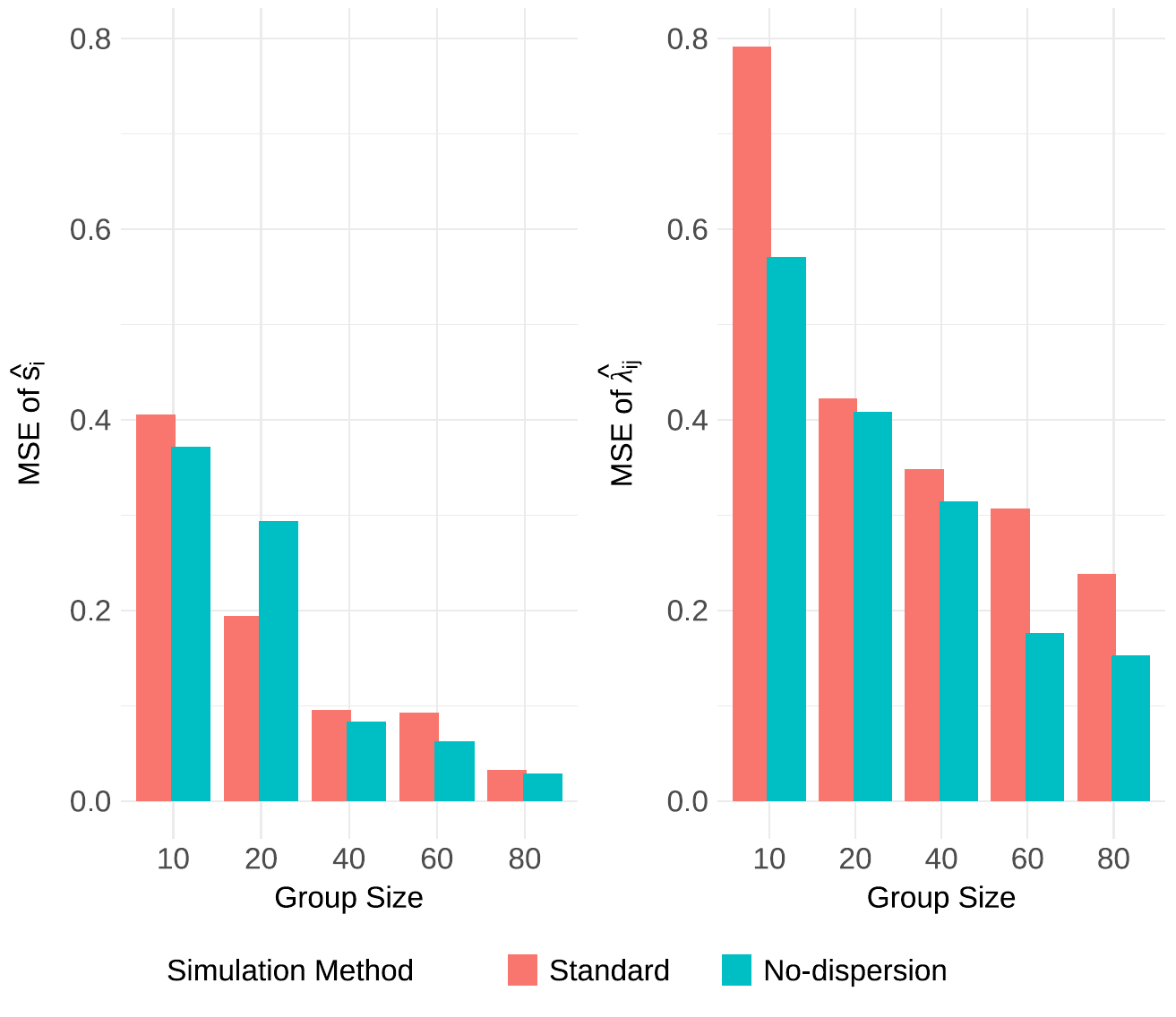}  
    \caption{
    Left panel: Mean squared error (MSE) of group--level estimates $\hat{s}_i$ across 1000 simulated datasets under two simulation setups: standard and no--dispersion. 
    Right panel: MSE of AE--level relative reporting rate estimates $\hat{\lambda}_{ij}$ under the same setups.
    }
    \label{fig:mse_sim_I}
\end{figure}

Performance was evaluated via mean squared error (MSE) at AE and group levels. Left Panel of Figure~\ref{fig:mse_sim_I} shows that the zGPS.AO method accurately estimates the group--level parameters \(s_i\), with larger AE groups yielding lower MSE due to greater information sharing across AEs. The results also demonstrate that the method performs well when applied to data without over--dispersion, where the ZINB model reduces to the ZIP model. Right Panel of Figure~\ref{fig:mse_sim_I} presents the AE--level results for \(\hat{\lambda}_{ij}\), showing similarly accurate estimation and robustness to the absence of over-dispersion. As expected, AE--level estimates exhibit higher MSE than group--level estimates, reflecting the smaller amount of information available for each individual AE.

\subsection{Simulation II}
\label{sec:simulation-study}
We conducted an additional simulation study to evaluate the performance of the zGPS.AO model in comparison with the simpler GPS model.
Count data were generated by combining $I=12$ drugs and $J=175$ adverse events.

The 175 AEs were further grouped into $K=6$ groups in accordance with the MedDRA hierarchy. The AE--level relative reporting rates $\lambda_{ij}$ were generated from Equation~\eqref{eq:lambda} with $p_i \sim \text{Unif}(0, 0.6)$, $\mu_i \sim Ga(5, 0.2)$ and $r = 5$. The counts $n_{ij}$ were generated from model~\eqref{eq:n:poi} the $E_{ij}$ parameters were tuned so as to mimick the Veneto drug safety data of Section~\ref{sec:results}.  This yielded datasets have on average 83\% of zeros, in accordance with the real dataset.    
We fitted the zGPS.AO and GPS models to 1,000 simulated datasets to compare AE--level relative reporting ratio estimates \(\hat{\lambda}_{ij}\). Unlike zGPS.AO, GPS does not incorporate the AE ontology.

\begin{figure}[t]
  \centering
  \begin{subfigure}[c]{0.45\textwidth}
    \centering
    \caption*{\normalsize (a)}

    \includegraphics[width=\textwidth]{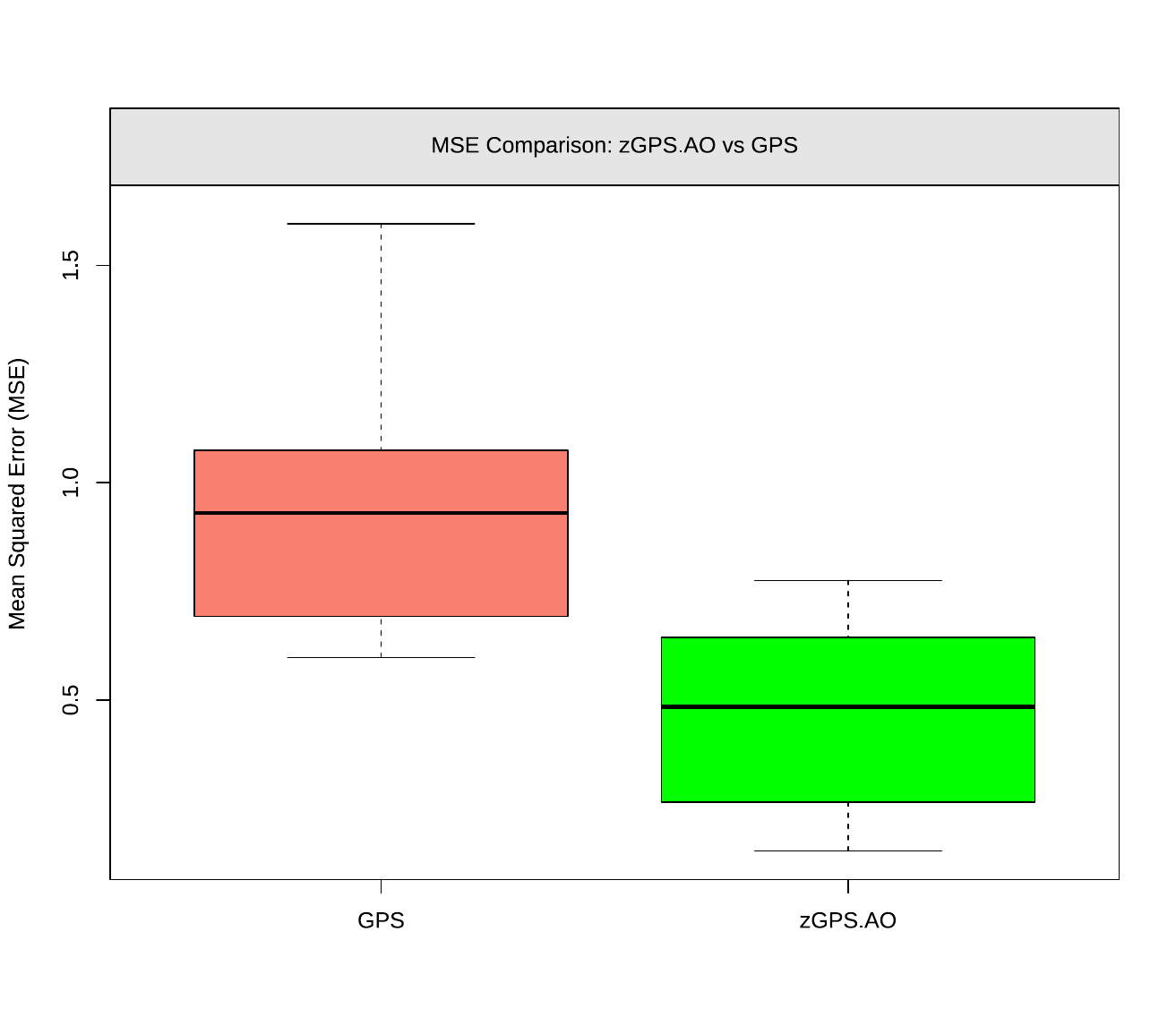}
  \end{subfigure}
  \hfill
  \begin{subfigure}[c]{0.45\textwidth}
    \centering
    \caption*{\normalsize (b)}

    \includegraphics[width=\textwidth]{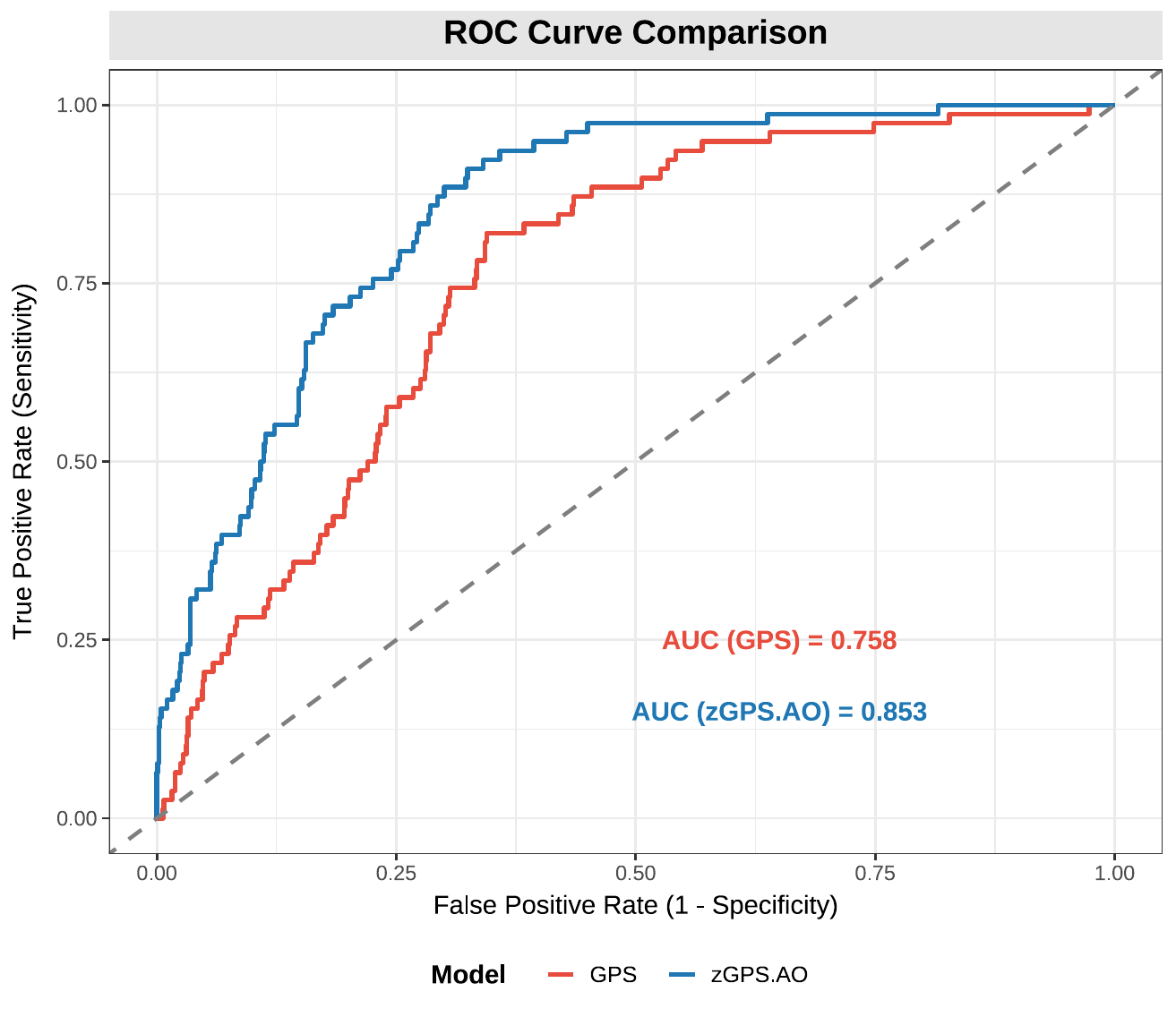}
  \end{subfigure}

  \caption{Summaries of numerical assessment of zGPS.AO versus GPS model fits. (a) Empirical distribution of AE--level MSE for \(\hat{\lambda}_{ij}\). (b) ROC curve for AE--level signal detection using \(\hat{\lambda}_{ij}\).}
  \label{fig:combined}
\end{figure}

Comparisons focused on individual AEs, as GPS cannot estimate group--level parameters. To evaluate the accuracy of parameter estimation and the performance of signal detection, we calculated the mean square error (MSE) and the area under the ROC curve (AUC), respectively. The MSE was defined as the squared difference between the estimator \(\hat{\lambda}_{ij}\) and the true value \(\lambda_{ij}\), averaged across all drug--AE pairs. For the AUC, an AE \(j\) for drug \(i\) was considered a true signal if \(\lambda_{ij} > 2\), and ROC curves were constructed using the corresponding \(\hat{\lambda}_{ij}\).
Panels (a) and (b) of Figure~\ref{fig:combined} summarize the findings of our numerical study. The zGPS.AO model clearly outperforms the GPS model, as indicated by a smaller MSE and larger AUC.  This improvement derives from sharing information between similar AEs in addition to the better handling of zero counts.

\subsection{Veneto drug safety data}
\label{sec:results}
We then applied the statistical methods discussed previously to a dataset of adverse event reports from 2023 to 2024 provided by the Pharmacovigilance center of the Italian Veneto region (CRFV) for $I=12$ drugs with hight reporting frequency.  The AEs were mapped to MedDRA's \textit{Preferred Terms} level, and AE groups where defined at the \textit{High Level Group Terms} level.  We excluded AEs with fewer than 15 occurrences and AE groups with fewer than 15 adverse events, leading to a total of 10,044 reports, $J=175$ adverse events and $K=6$ AE groups.  To compute drug--AE counts, we applied a weighting strategy.  Single-drug reports were assigned a weight of 1, while multi--drug reports assigned each drug--AE pair a weight which was inversely proportional to the number of drugs mentioned, assuming that the AE is linked to each drug with equal probability.  Final counts were obtained by adding these weights and rounding to the nearest integer.

We fitted both the zGPS.AO and GPS models to the final dataset, achieving a Pearson correlation coefficient of 0.64 between their AE--level relative reporting rates (RR) estimates.  In the zGPS.AO model, an AE was classified as a safety signal if its AE--level RR estimate exceeded 2 and the q-value was less than 0.05.  Conversely, for the GPS model, the lower bound of the 99\% confidence interval for the AE--level reporting ratio had to exceed 2.  Based on these criteria, the zGPS.AO model identified 245 AE signals across the 12 studied drugs, which include the only 15 signals detected by the GPS model.

\begin{figure}[ht]
    \centering
    \begin{subfigure}{0.45\textwidth}
    \centering
    \caption*{\normalsize (a)}
    \includegraphics[width=\linewidth]{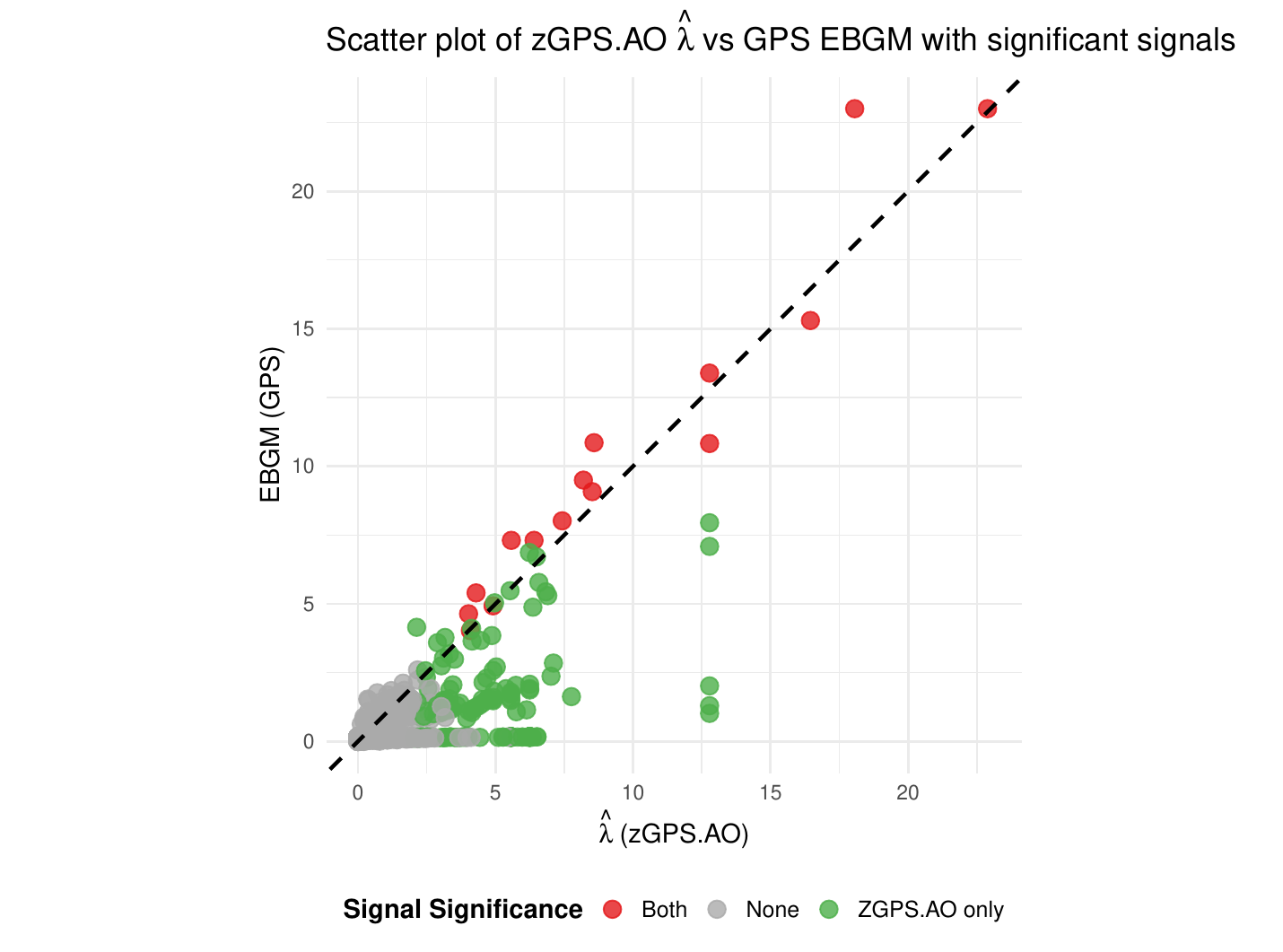}
        \label{fig:fig1}
    \end{subfigure}
    \hfill
    \begin{subfigure}{0.45\textwidth}
        \centering
        \caption*{\normalsize (b)}
    \includegraphics[width=\linewidth]{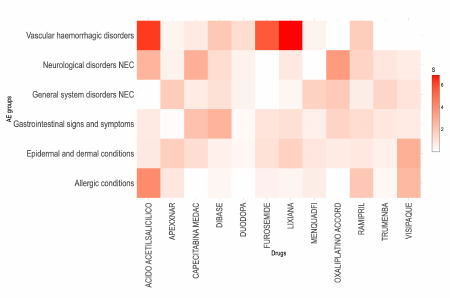}
        \label{fig:fig2}
    \end{subfigure}
    \caption{
    Results from real data analysis comparing zGPS.AO and GPS models. 
    (a) Scatter plot of AE--level relative reporting rates estimated by the two methods. 
    Grey points indicate signals detected by neither method, green points indicate signals detected only by zGPS.AO and red points indicate signals detected by both methods. The correlation between RR estimates is \( r = 0.64 \). 
    (b) Heatmap showing group--level relative risks for selected adverse event (AE) groups (rows) and drugs (columns). The more red the color, the stronger are the associations between drug and AE group.
    }
  \label{fig:combined_II}
\end{figure}

Plotting the group--level RRs, $\hat s_i^k$, from the zGPS.AO model fit for all drug--AE group combinations, as in panel (b) of Figure~\ref{fig:combined_II}, helps identifying potential drug safety issues. We define an AE group as a safety concern when its $q$-value \(< 0.05\) and $\hat s^k > 2$. To illustrate, strong associations were observed for \textit{acido acetilsalicilico} with \textit{Neurological disorders NEC} and \textit{Allergic conditions}, \textit{capecitabina medac} and \textit{oxaliplatino accord} with \textit{Neurological disorders NEC}, \textit{furosemide} and \textit{lixiana} with \textit{Vascular haemorrhagic disorders}, and \textit{visipaque} with \textit{Allergic conditions} and \textit{Epidermal and dermal conditions}. A list of the constituent AEs for each highlighted drug--AE group is provided in Figure~\ref{fig:all_images}.

\begin{sidewaysfigure}
\centering
\begin{subfigure}[b]{0.45\textheight}
    \includegraphics[width=\textwidth,height=0.32\textheight]{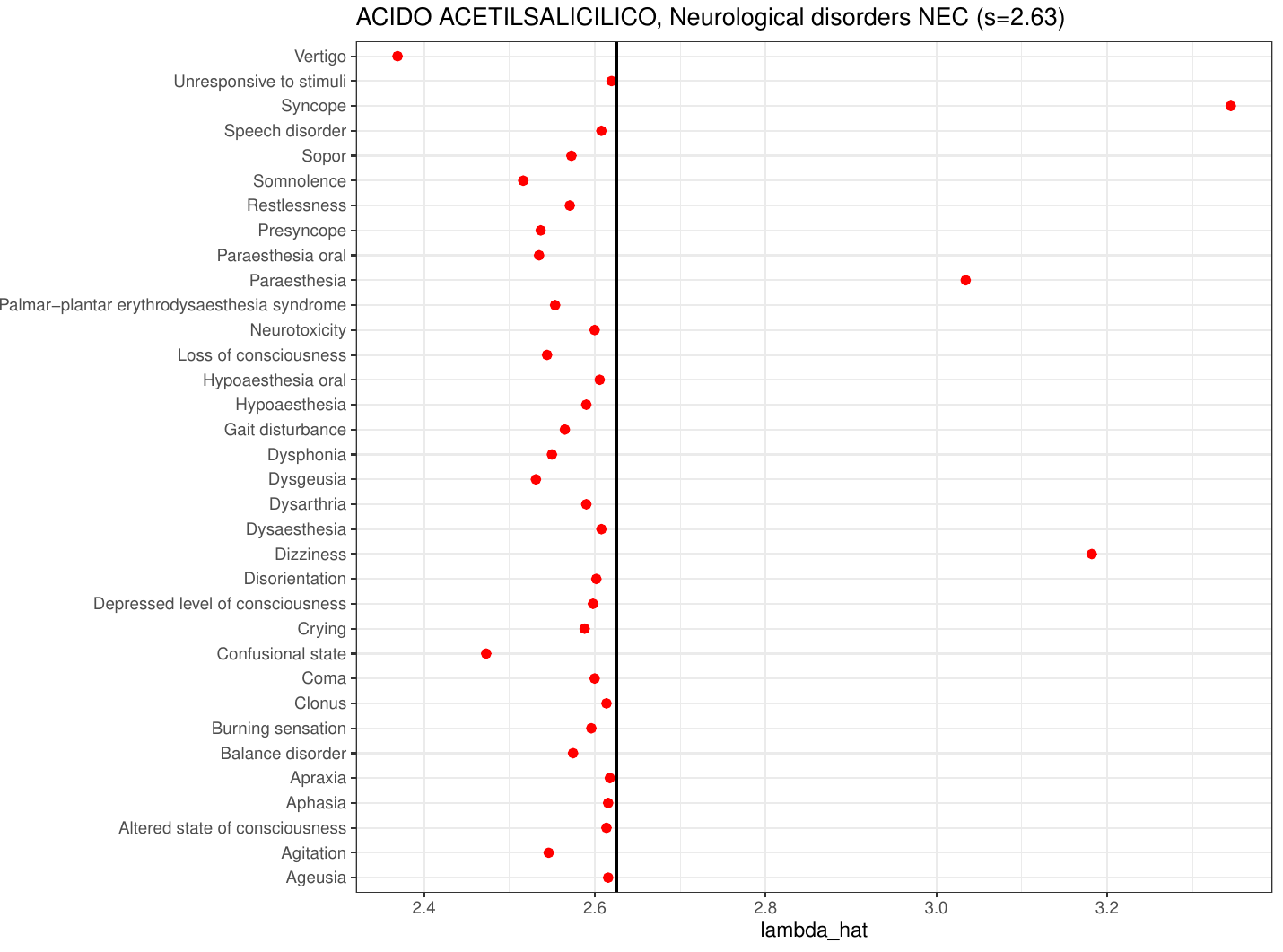}
    \caption{}
\end{subfigure}
\hspace{0.05\textheight}
\begin{subfigure}[b]{0.45\textheight}
    \includegraphics[width=\textwidth,height=0.32\textheight]{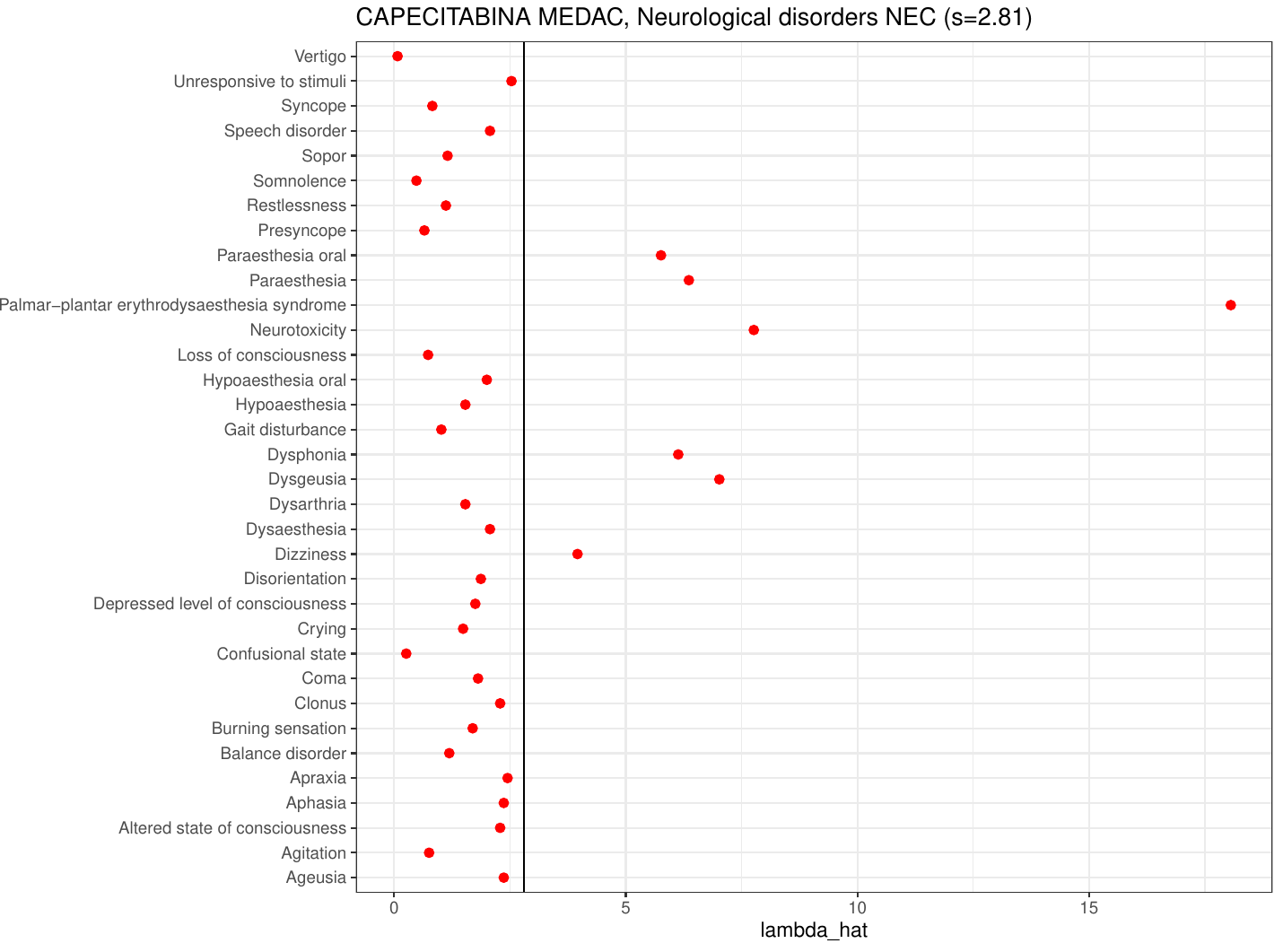}
    \caption{}
\end{subfigure}

\vspace{0.6cm}

\begin{subfigure}[b]{0.45\textheight}
    \includegraphics[width=\textwidth,height=0.32\textheight]{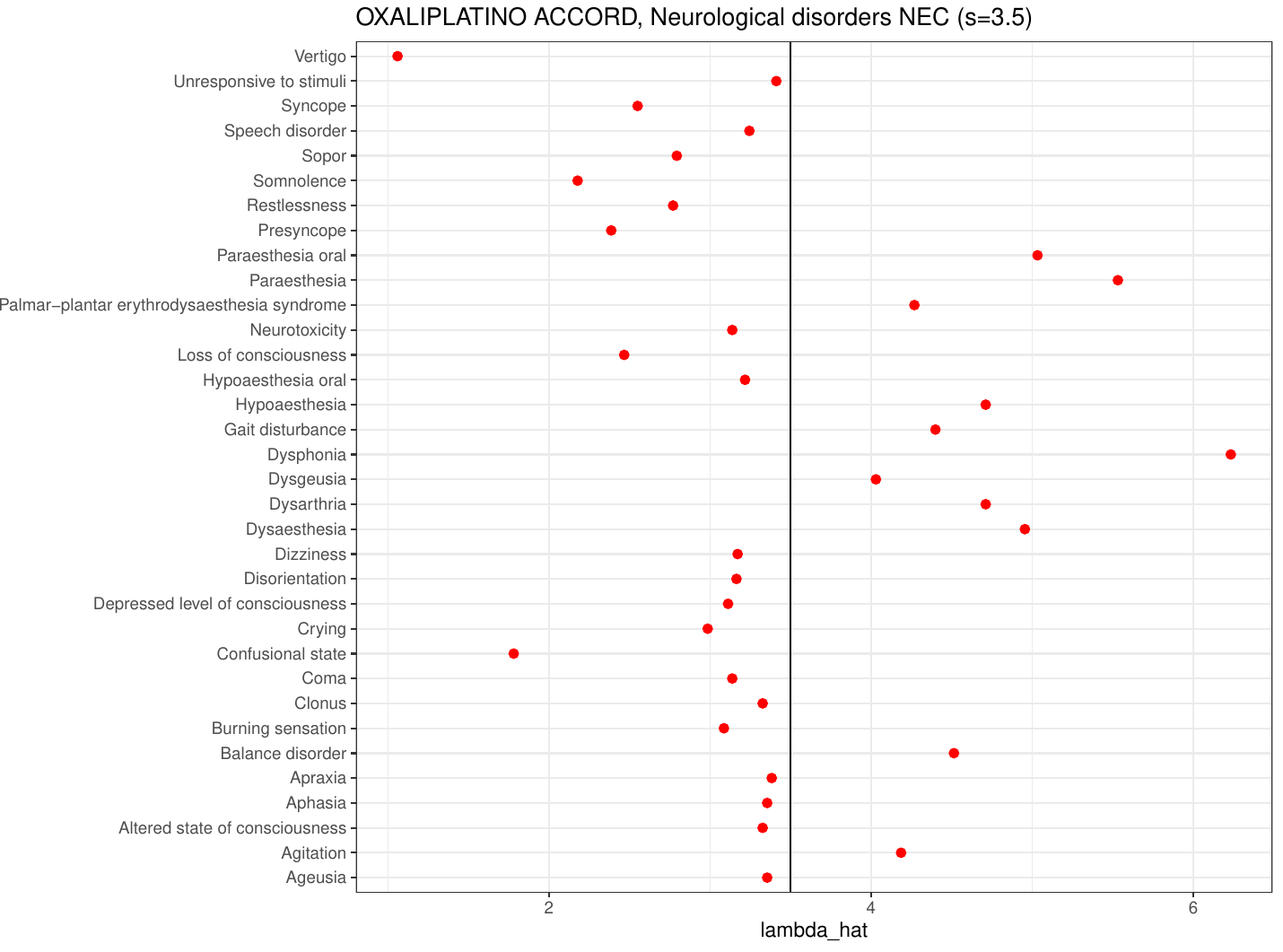}
    \caption{}
\end{subfigure}
\hspace{0.05\textheight}
\begin{subfigure}[b]{0.45\textheight}
    \includegraphics[width=\textwidth,height=0.32\textheight]{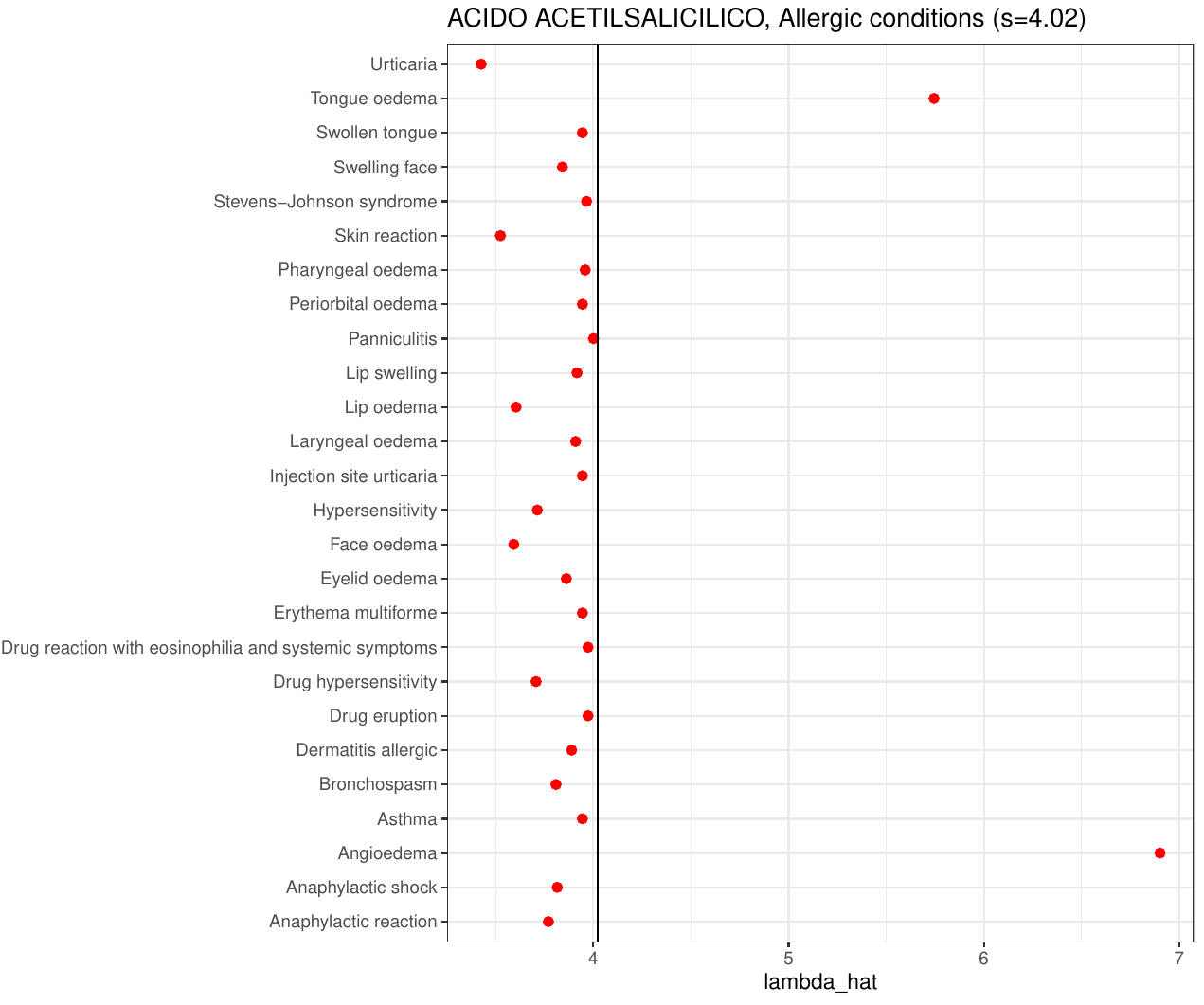}
    \caption{}
\end{subfigure}

\caption{
Adverse events for selected drugs across AE groups. Panels (a--c) show the \textit{Neurological disorders NEC} group, and panel (d) the \textit{Allergic conditions} group.
The $x$-axis represents estimated reporting ratios ($\hat{\lambda}_{ij}$), the $y$-axis lists specific AEs within each group, and the vertical line in each panel indicates the group--level relative risk (\(s\)) for the corresponding drug--AE group.}
\label{fig:all_images}
\end{sidewaysfigure}

\begin{sidewaysfigure}
\ContinuedFloat
\centering
\begin{subfigure}[b]{0.45\textheight}
    \includegraphics[width=\textwidth,height=0.32\textheight]{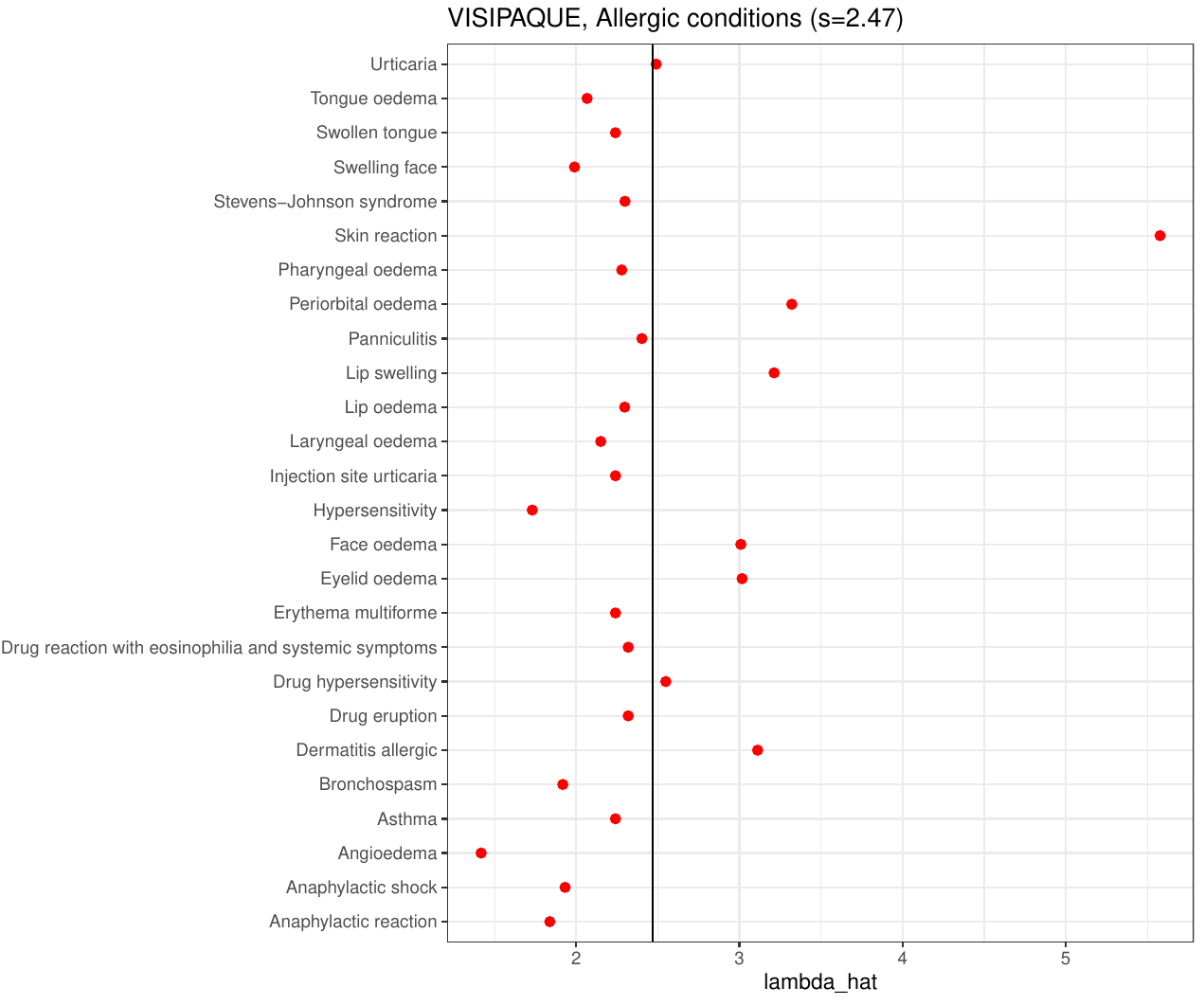}
    \caption{}
\end{subfigure}
\hspace{0.05\textheight}
\begin{subfigure}[b]{0.45\textheight}
    \includegraphics[width=\textwidth,height=0.32\textheight]{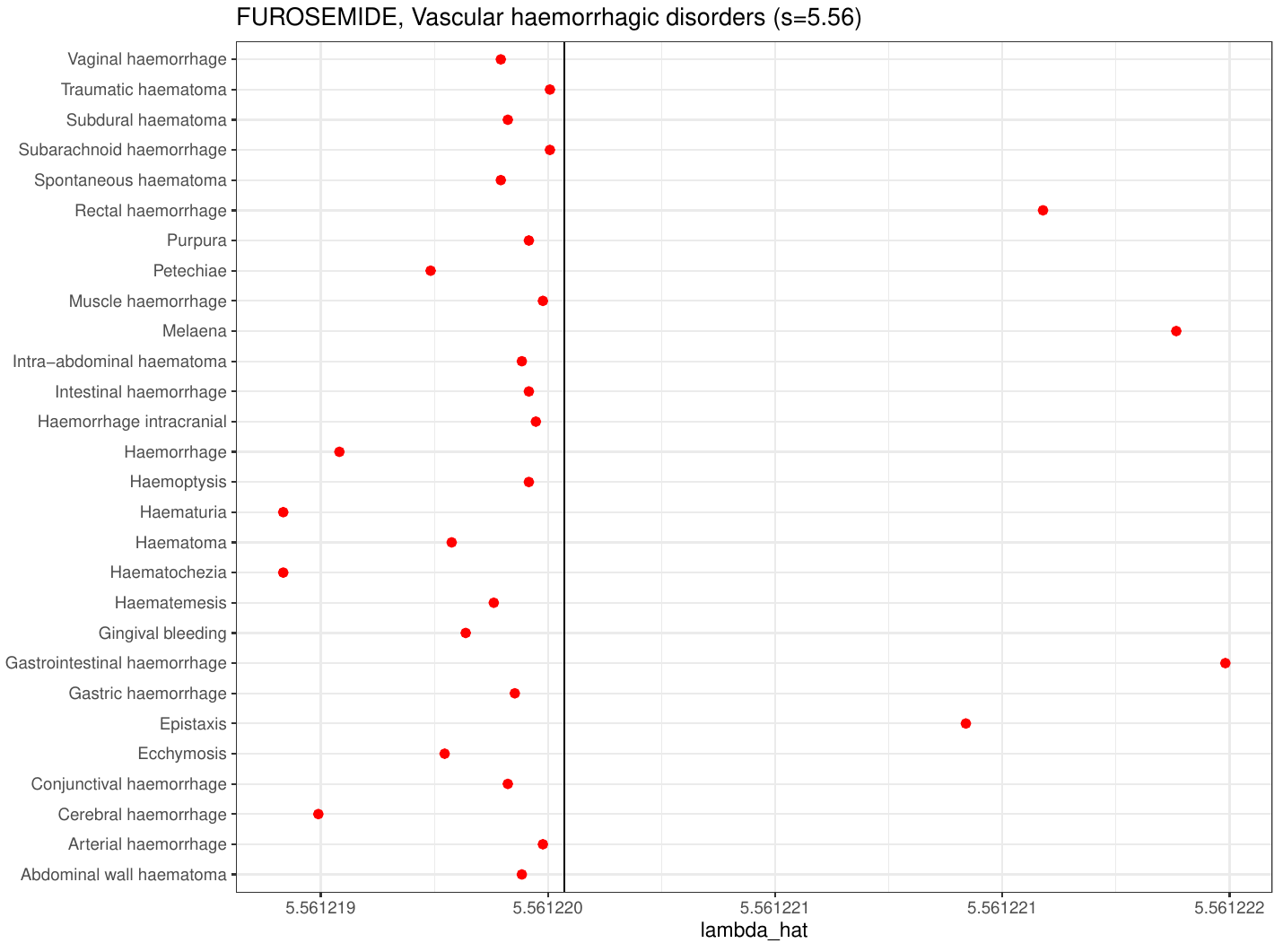}
    \caption{}
\end{subfigure}

\vspace{0.6cm}

\begin{subfigure}[b]{0.45\textheight}
    \includegraphics[width=\textwidth,height=0.32\textheight]{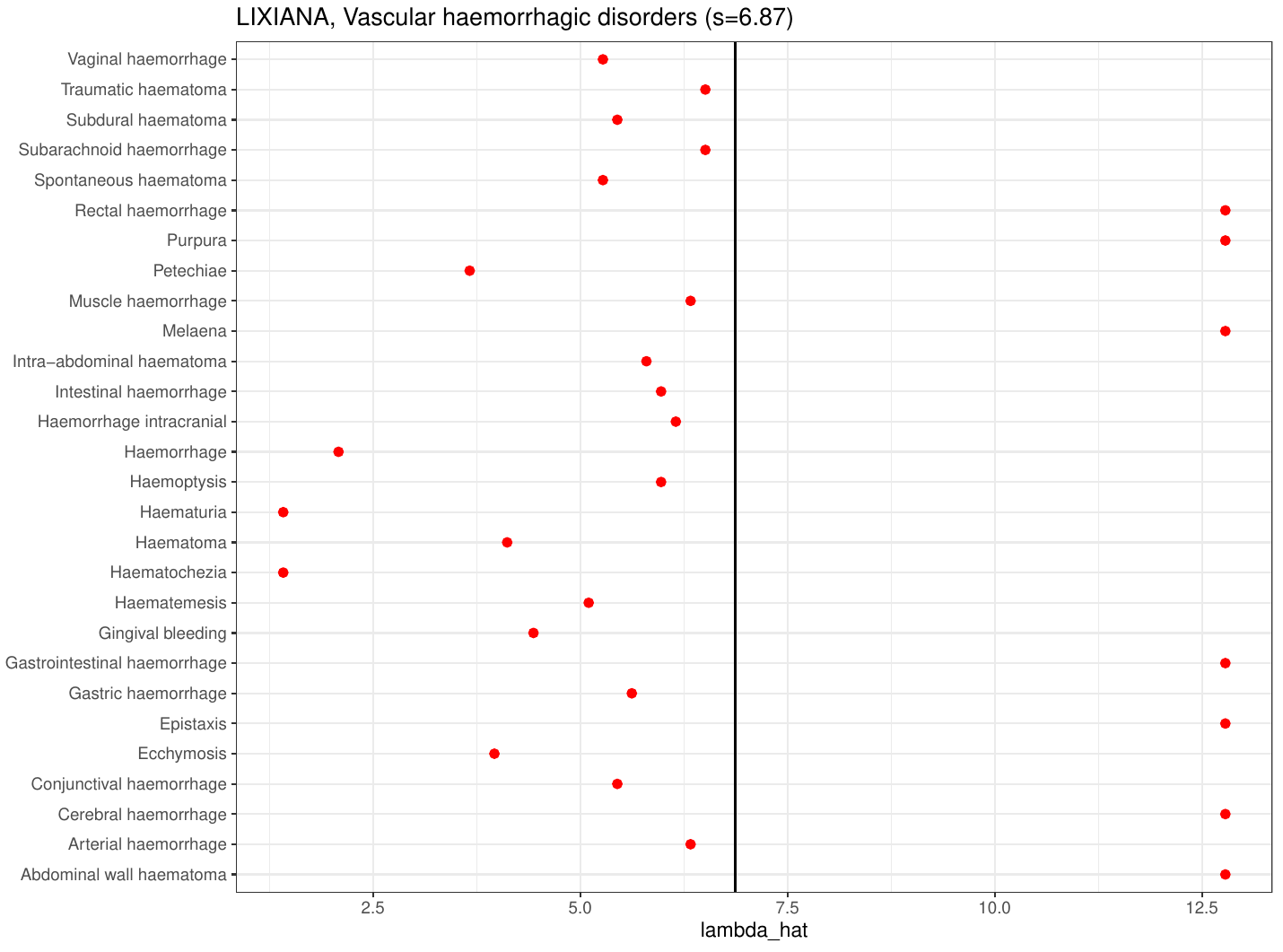}
    \caption{}
\end{subfigure}
\hspace{0.05\textheight}
\begin{subfigure}[b]{0.45\textheight}
    \includegraphics[width=\textwidth,height=0.32\textheight]{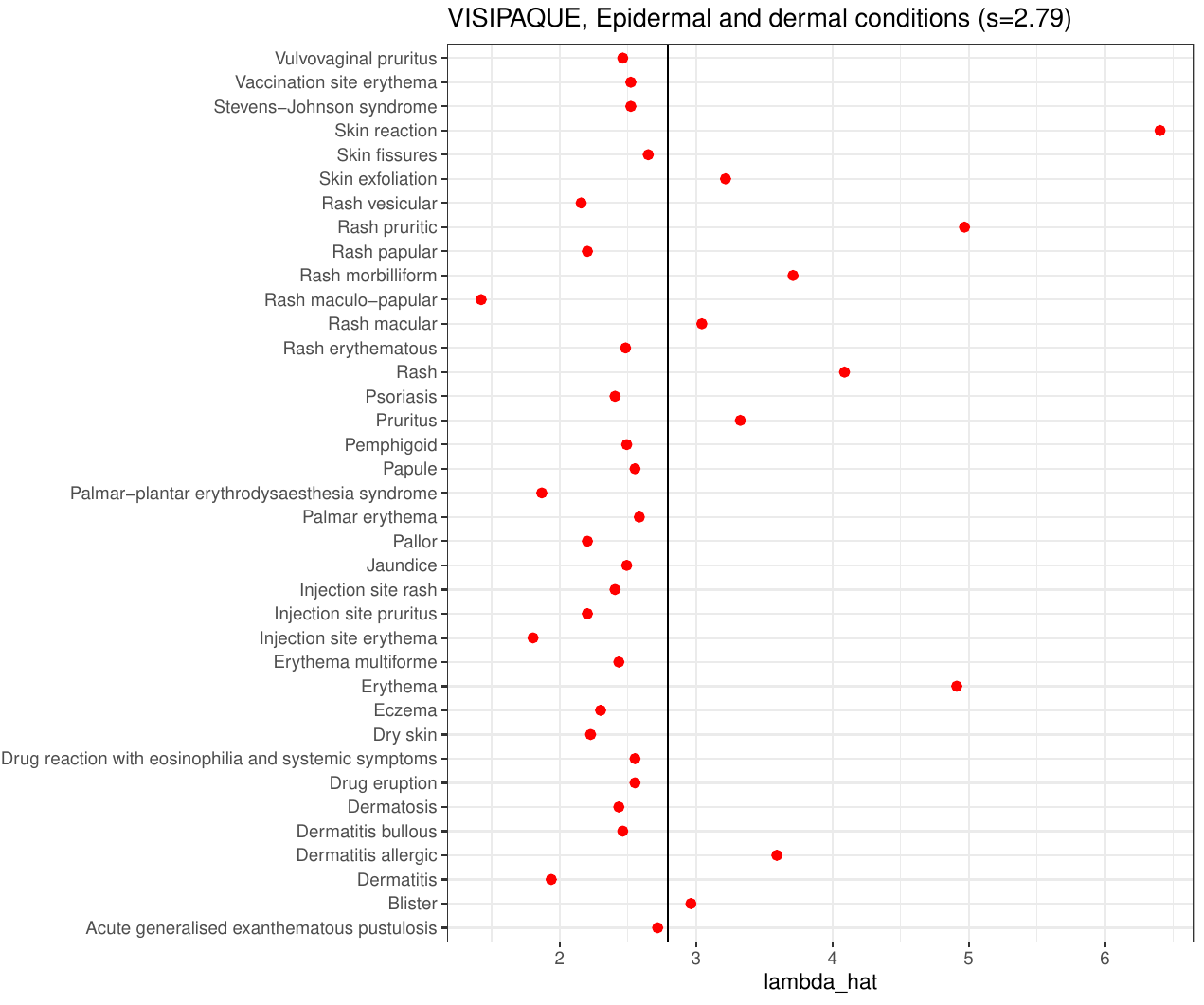}
    \caption{}
\end{subfigure}

\caption{Continued: Panels (e--g) show the \textit{Allergic} and \textit{Vascular haemorrhagic disorders} groups, and panel (h) the \textit{Epidermal and dermal conditions} group. The $x$-axis represents estimated reporting ratios ($\hat{\lambda}_{ij}$), the $y$-axis lists specific AEs within each group, and the vertical line in each panel indicates the group--level relative risk (\(s\)) for the corresponding drug--AE group.}
\label{fig:all_images_2}
\end{sidewaysfigure}

\section{Assessing model performance}
\label{sec:model_performance}
To better evaluate model performance, we compared three approaches for handling drug--AE counts: data thinning, random sample splitting and stratified sample splitting. In stratified sample splitting, all data points are preserved by first expanding aggregated counts into individual instances before dividing them into training and validation sets. Model performance was assessed using mean squared error (MSE), representing the average squared difference between observed and predicted counts, with smaller values indicating better fit. For classification, counts were labeled positive ($n_{ij} > 0$) or negative ($n_{ij} = 0$), and predicted values were converted into binary outcomes using a 0.5 threshold to compute AUC, accuracy, sensitivity, specificity, precision, and F1 score. For each method, MSE was calculated at training proportions $\varepsilon = 0.5, 0.7, 0.8,$ and $0.9$, with each setting repeated ten times and results reported as the average across repetitions.

\begin{sidewaysfigure}[tp]
    \centering  
         
    \begin{subfigure}[b]{0.32\textheight}         
        \includegraphics[width=\textwidth,height=0.3\textheight]{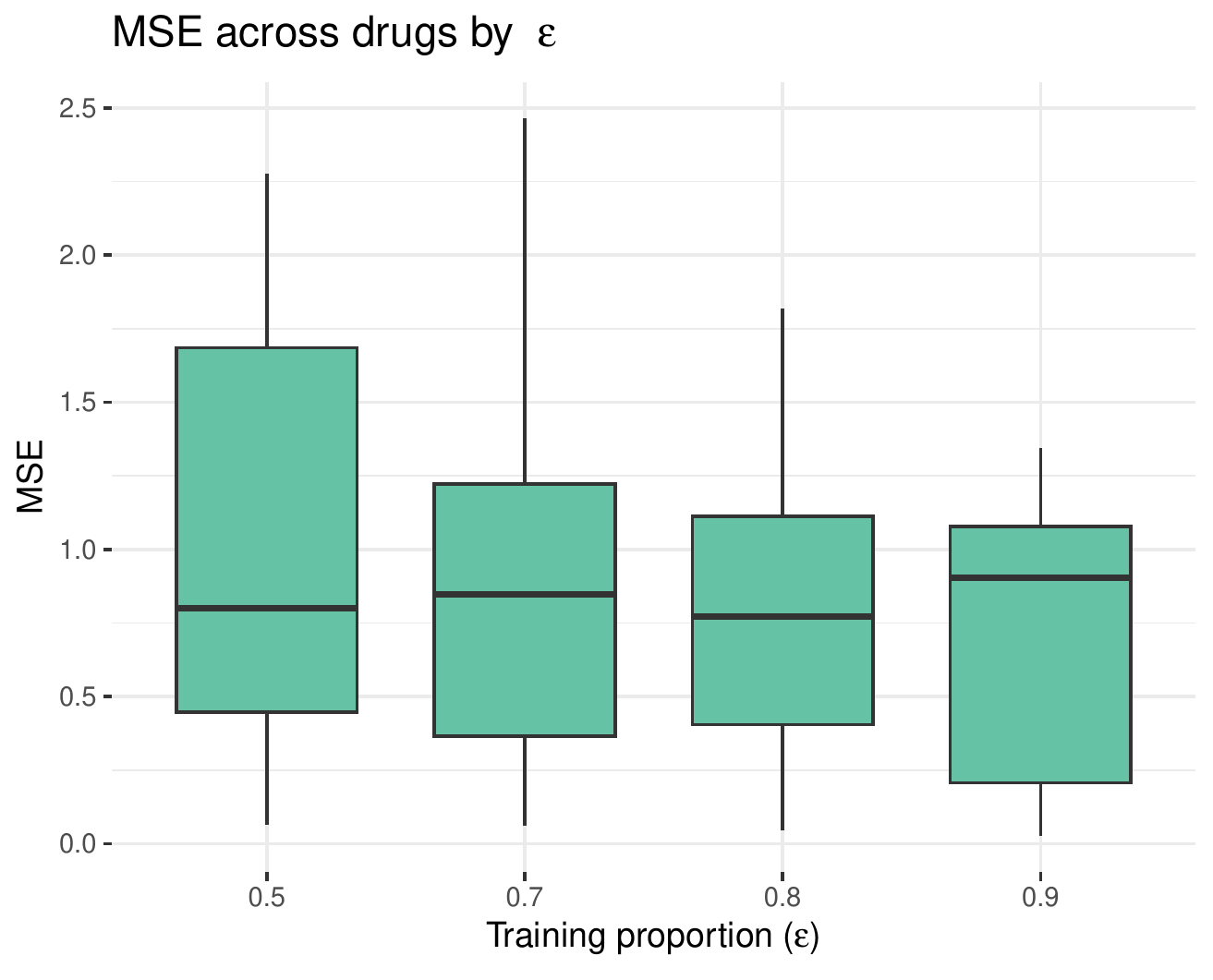}         
        \caption{}     
    \end{subfigure}     
    \begin{subfigure}[b]{0.32\textheight}         
        \includegraphics[width=\textwidth,height=0.3\textheight]{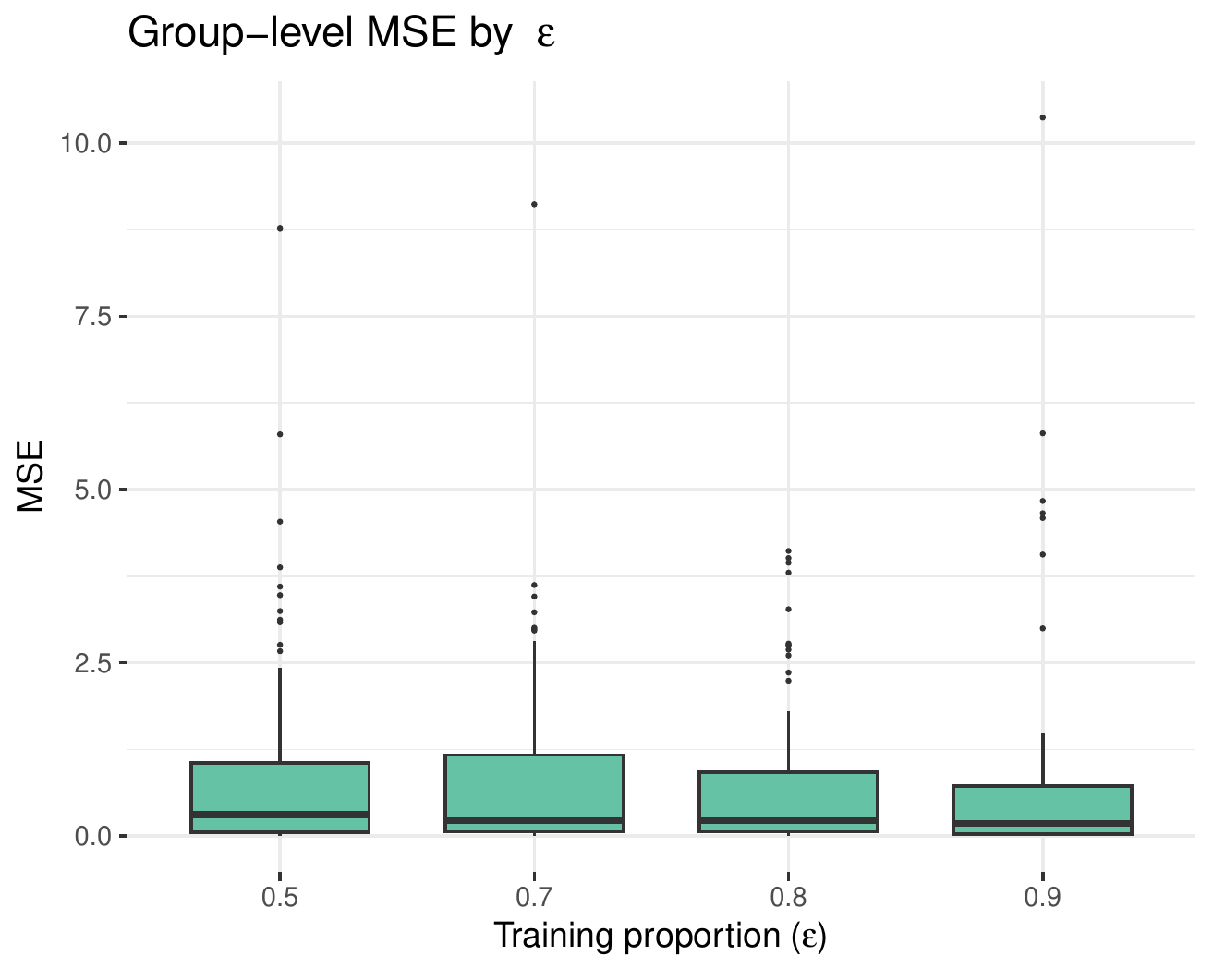}         
        \caption{}     
    \end{subfigure}     
    \begin{subfigure}[b]{0.32\textheight}         
        \includegraphics[width=\textwidth,height=0.3\textheight]{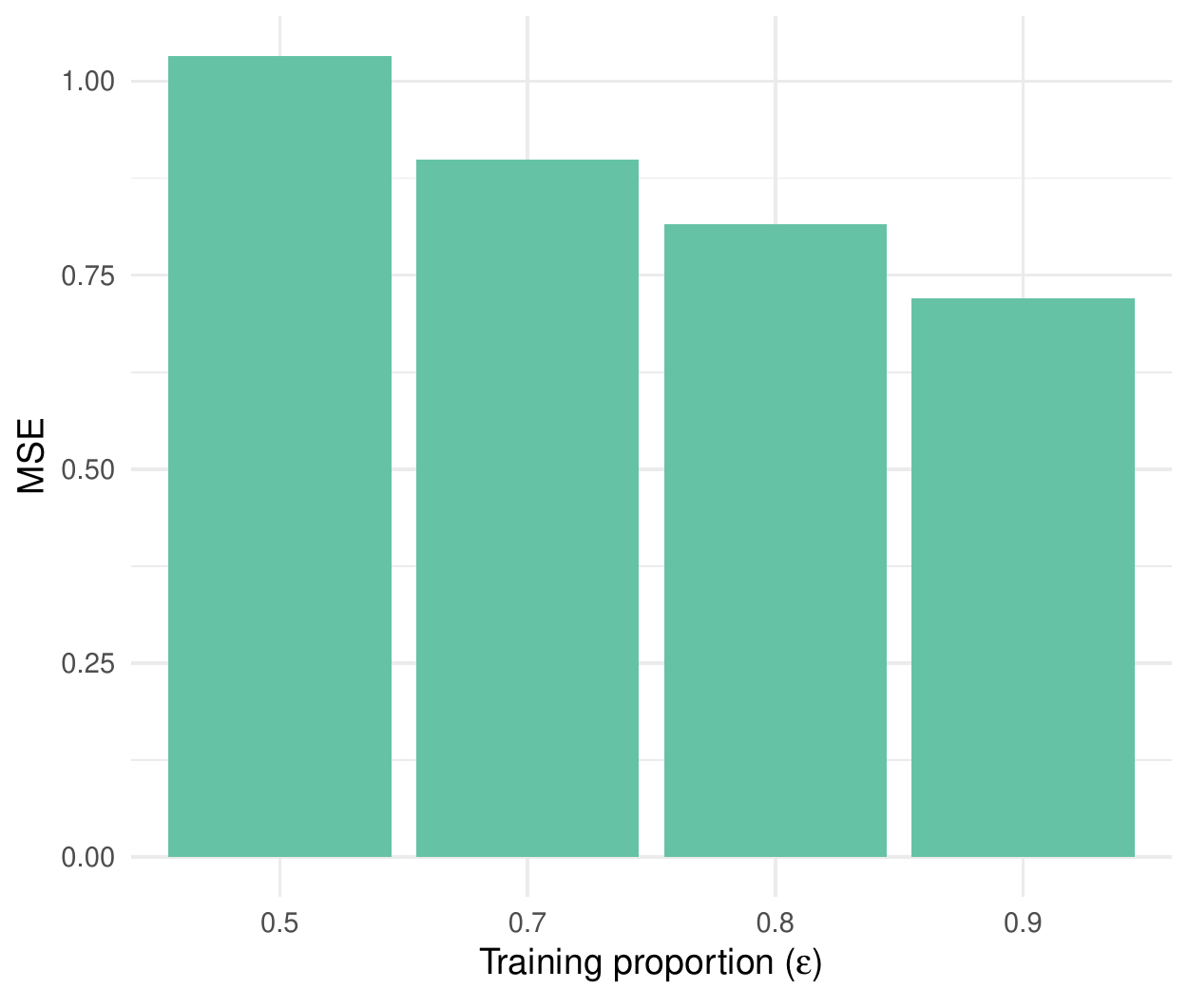}         
        \caption{}     
    \end{subfigure}

    \vspace{0.6cm}   
    \begin{subfigure}[b]{0.32\textheight}         
        \includegraphics[width=\textwidth,height=0.3\textheight]{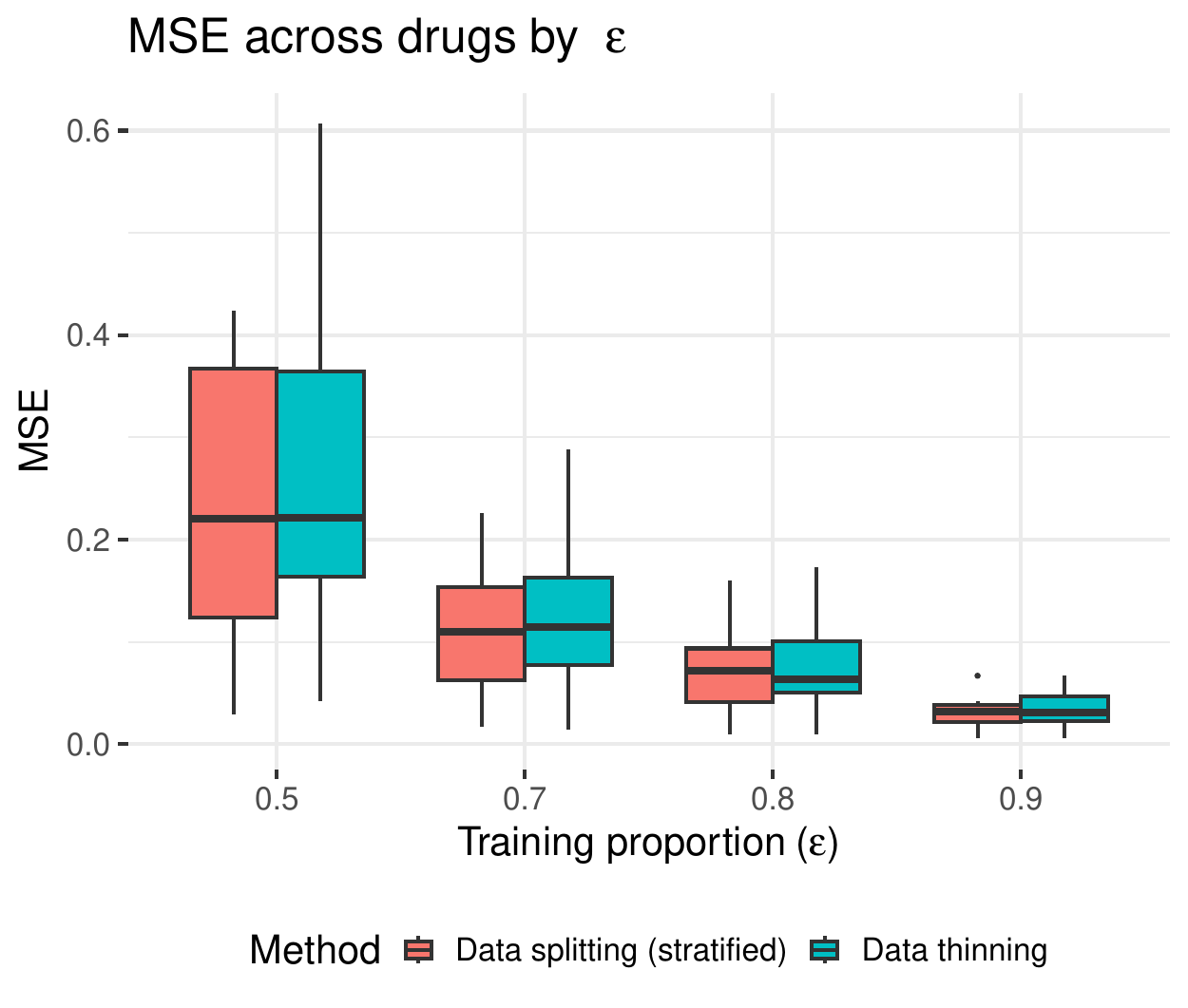}         
        \caption{}     
    \end{subfigure}     
    \begin{subfigure}[b]{0.32\textheight}         
        \includegraphics[width=\textwidth,height=0.3\textheight]{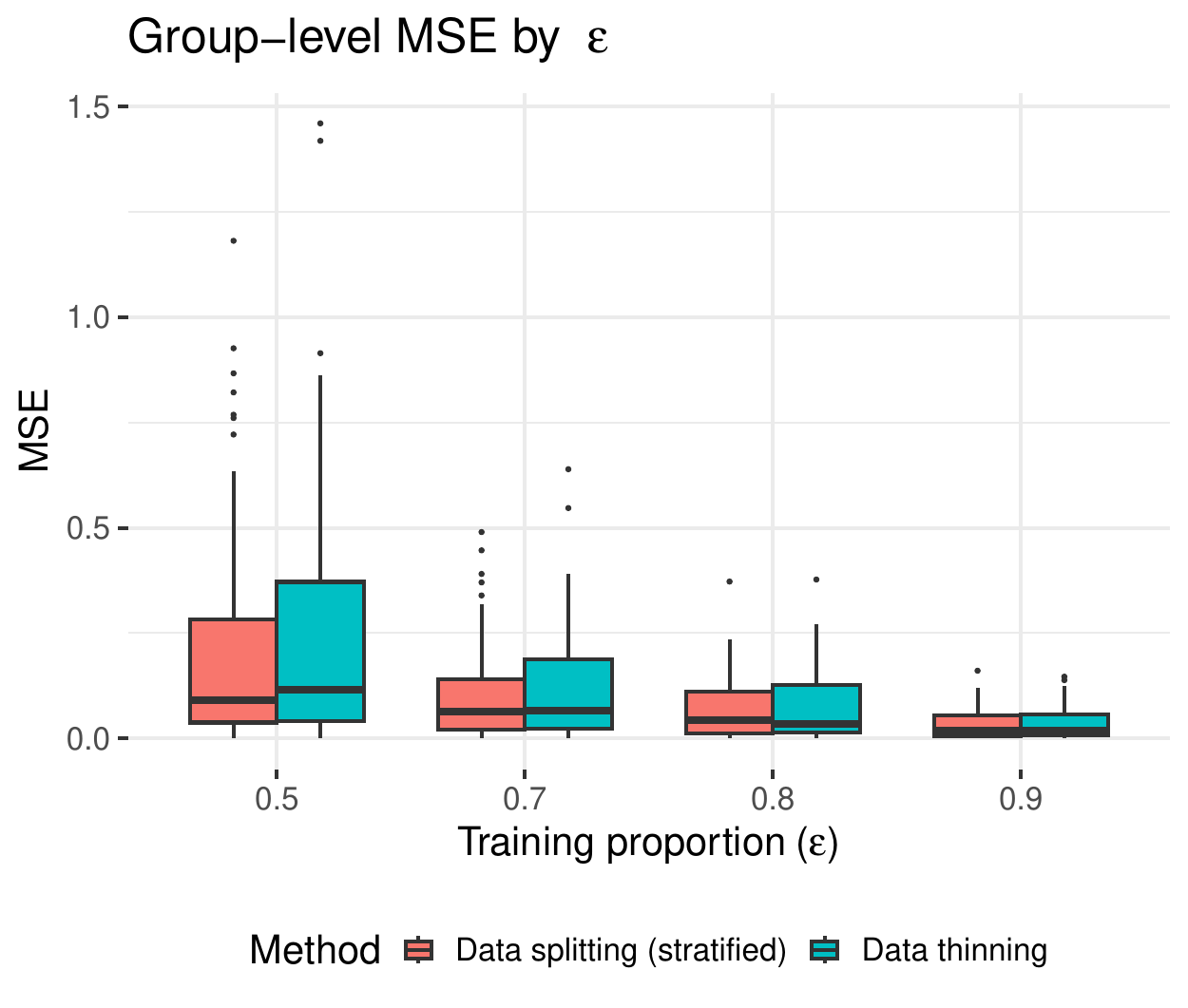}        
        \caption{}  
        \label{fig:group_level} 
    \end{subfigure}     
    \begin{subfigure}[b]{0.32\textheight}         
        \includegraphics[width=\textwidth,height=0.3\textheight]{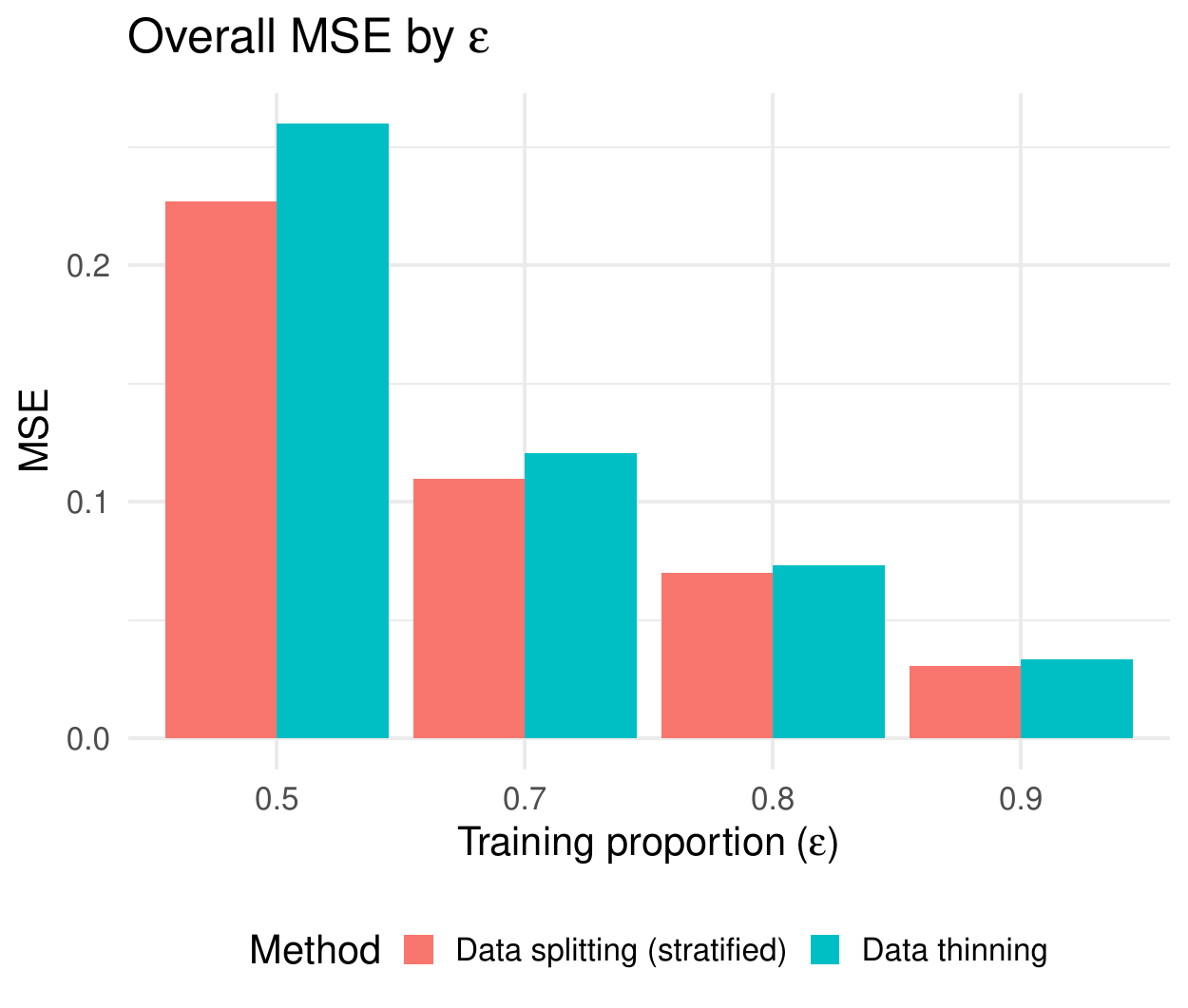}   
        \caption{}     
    \end{subfigure}      

    \caption{
        Mean squared error (MSE) of predicted drug--AE counts compared to observed counts, across training proportions ($\varepsilon$) for three data partitioning strategies in the drug--adverse event dataset. Panels (a--c) show random splitting, and panels (d--f) show stratified splitting and the proposed data thinning method.}
    \label{fig:mse_comparison_all} 
\end{sidewaysfigure}

Figure~\ref{fig:mse_comparison_all} compares mean squared error (MSE) across training fractions for the three strategies: random sample splitting, stratified splitting, and the proposed data thinning method.
With random sample splitting, certain AEs were randomly absent from the validation set, resulting in the exclusion of all corresponding drug--AE pairs and highlighting that random splitting can undermine model validation in a pharmacovigilance setting.
In contrast, stratified splitting and the proposed data thinning method preserved the full set of drug--AE pairs in both training and validation, avoiding the coverage loss observed with random splitting. As shown in the group--level MSE plot (Figure \ref{fig:group_level}), both stratified splitting and data thinning methods maintained low variability across training fractions, with median MSE decreasing as the training fraction increased. Although data thinning achieved slightly higher median MSE at lower fractions (e.g., $\varepsilon=0.5$), it exhibited fewer extreme outliers than stratified splitting. This suggests more stable performance in capturing group--level AE patterns while ensuring complete representation of drug--AE pairs.

\begin{figure}[tp]
    \centering
\includegraphics[width=0.8\textwidth]{Figures/Overall_Metric.pdf}  
       \caption{Model performance across different data partitioning strategies over different training proportions ($\varepsilon = 0.5, 0.7, 0.8, 0.9$). Partitioning strategies: data thinning (blue), stratified splitting (red) and random splitting (green). Metrics include accuracy, AUC, F1 score, precision, sensitivity and specificity. Stratified splitting provides the most balanced performance across metrics; data thinning performs well at moderate $\varepsilon$, while random splitting is less reliable overall.}

    \label{fig:metric_comparison_all}
\end{figure}
Figure~\ref{fig:metric_comparison_all} and Table~\ref{tab:results} summarize the comparative performance of data thinning, stratified splitting and random splitting. Stratified splitting consistently achieves the highest accuracy and AUC, with strong specificity, making it the most balanced strategy. Data thinning approaches stratified performance in accuracy and AUC but shows declining sensitivity and F1 at higher \(\varepsilon\), reflecting reduced detection of true positives. Random splitting provides stable sensitivity and relatively consistent precision but underperforms in specificity and overall reliability. Overall, stratified splitting offers the best balance across metrics, data thinning performs well at moderate \(\varepsilon\) and random splitting is less robust despite its sensitivity advantage.

\begin{table}[tp]
\centering
\caption{Comparison of performance metrics: AUC, accuracy, sensitivity, specificity, precision and F1 score-for  data thinning (Th), stratified data splitting (Ss), and random data splitting (Rs)  across different training proportions \(\varepsilon\).}
\label{tab:results}
\footnotesize
\setlength{\tabcolsep}{2pt}
\begin{tabular}{@{} l *{3}{S[table-format=1.3]} *{3}{S[table-format=1.3]} *{3}{S[table-format=1.3]} *{3}{S[table-format=1.3]} *{3}{S[table-format=1.3]} *{3}{S[table-format=1.3]} @{}}
\toprule
{$\varepsilon$} & \multicolumn{3}{c}{AUC} & \multicolumn{3}{c}{Accuracy} & \multicolumn{3}{c}{Sensitivity} & \multicolumn{3}{c}{Specificity} & \multicolumn{3}{c}{Precision} & \multicolumn{3}{c}{F1} \\
\cmidrule(lr){2-4} \cmidrule(lr){5-7} \cmidrule(lr){8-10} \cmidrule(lr){11-13} \cmidrule(lr){14-16} \cmidrule(lr){17-19}
 & {Th} & {Ss} & {Rs} & {Th} & {Ss} & {Rs} & {Th} & {Ss} & {Rs} & {Th} & {Ss} & {Rs} & {Th} & {Ss} & {Rs} & {Th} & {Ss} & {Rs} \\
\midrule
0.5 & 0.879 & 0.908 & 0.814   & 0.919 & 0.926 & 0.879 & 0.460 & 0.521 & 0.480 & 0.961 & 0.962 & 0.937 & 0.518 & 0.555 & 0.526 & 0.487 & 0.537 &0.499  \\
0.7 & 0.915 & 0.932 & 0.850 & 0.945 & 0.945 & 0.883 & 0.439 & 0.447 & 0.510 & 0.976 & 0.977 & 0.937 & 0.525 & 0.548 & 0.533 & 0.477 & 0.492 &0.520  \\
0.8 & 0.915 & 0.943 & 0.865 & 0.957 & 0.957 & 0.880 & 0.333 & 0.385 & 0.511 & 0.984 & 0.984 & 0.935 & 0.488 & 0.528 & 0.534 & 0.395 & 0.444 &0.520  \\
0.9 & 0.917 & 0.960 & 0.868 & 0.974 & 0.973 &0.883  & 0.118 & 0.239 & 0.485 & 0.996 & 0.994 & 0.940 & 0.450 & 0.509 &0.536  & 0.185 & 0.323 &0.503  \\
\bottomrule
\end{tabular}
\end{table} 

\begin{figure}[tp]
    \centering
    \includegraphics[width=0.9\textwidth]{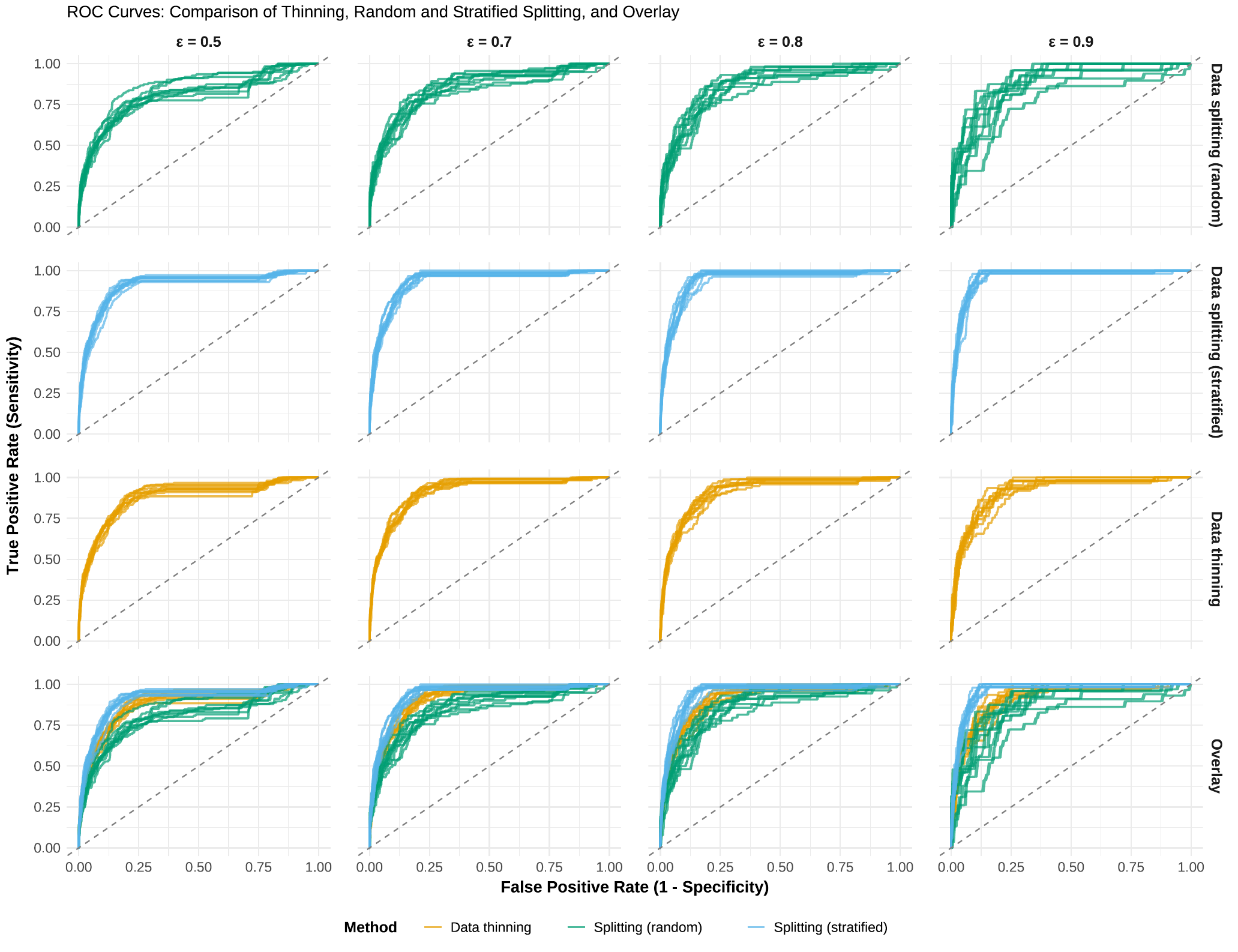}
    \caption{    ROC curves for random splitting, stratified splitting and data thinning across training proportions ($\varepsilon = 0.5, 0.7, 0.8, 0.9$). Each method was repeated 10 times with different random seeds (123--132). Stratified splitting and data thinning yield more stable and higher curves than random splitting, which suffers from variability due to the loss of rare drug–AE pairs. The diagonal dashed line denotes the baseline AUC of 0.5.}
    \label{fig:Roc_all}
\end{figure}
Figure~\ref{fig:Roc_all} compares the ROC performance of the three partitioning strategies across training proportions ($\varepsilon = 0.5, 0.7, 0.8, 0.9$). Stratified splitting and data thinning produced consistently stable ROC curves across ten different replications, while random splitting showed greater variability. These results indicate that stratified splitting and data thinning provide more robust and interpretable model evaluation.

\section{Discussion}
\label{sec:applications_discussion}
\noindent
Our findings show that the zGPS.AO model enhances ADR signal detection by combining AE ontology with a zero--inflated negative binomial (ZINB) framework while addressing excess zeros and correlations among AEs. Simulation studies across 1,000 datasets showed that zGPS.AO achieves lower mean squared error (MSE) in AE--level relative reporting ratio estimates (\(\hat{\lambda}_{ij}\)) and higher area under the curve (AUC) for AE--level signal detection than GPS, benefiting from information sharing across related AEs and better accommodation of sparse data. Application to the Veneto pharmacovigilance database confirmed these advantages: zGPS.AO identified more signals, including all those detected by GPS, while improving discrimination of true drug--AE associations and reducing false negatives.  

Furthermore, we introduced a data--thinning strategy for post--selection inference in pharmacovigilance statistical modeling. Unlike conventional random splitting, which may omit  drug--AE pairs, data thinning exploits the convolution--closed property of the Poisson distribution to generate independent training and validation sets without losing any drug--AE information. Comparative evaluation showed that data thinning provides stable and reliable model performance, approaching the effectiveness of stratified splitting while avoiding biases introduced by random partitioning.  

Despite these advances, the current framework focuses solely on drug--AE pairs and does not incorporate patient--level covariates such as demographics, comorbidities, or concomitant medications, which are critical for distinguishing causal drug--AE relationships from confounding. Future extensions will integrate these variables, as well as drug interaction networks and treatment parameters, to enhance signal detection and post--market pharmacovigilance monitoring. Analyses were performed in R version 4.4.2 (R Core Team, 2024) on a MacBook Pro (1.4 GHz Quad-Core Intel Core i5, 16 GB RAM). Generating 4000 permutations, implemented using parallel processing, required approximately 1 hr 40 min, with an additional 1 min for parameter estimation. Simulation I (1000 runs) required 22 min, Simulation II (1000 runs) required 15 min, and model validation 5 min. To support reproducibility, all scripts used to generate the results reported in this paper are publicly available at \url{https://github.com/kenenitadesse/zGPS-AO-PSI}, with detailed descriptions of the data format to allow researchers with similar data to reproduce the analyses. The original Veneto pharmacovigilance data used in this work are owned by the Regional Center for Pharmacovigilance and cannot be publicly shared.

\section*{Acknowledgments}
The publication was produced with funding from the Italian Ministry of University and Research under the Call for Proposals related to the scrolling of the final rankings of the PRIN 2022 call -- Project title \virgolette{Latent variable models and dimensionality reduction methods for complex data} -- Project No. 20224CRB9E, CUP C53C24000730006 -- PI Prof. Paolo Giordani -- Grant Assignment Decree No. 1401 adopted on 18.9.2024 by MUR.

\bibliographystyle{unsrtnat}
\bibliography{references}

\clearpage

\appendix

\section*{Appendix}

\section{Negative binomial and zero-inflated negative binomial models}\label{app:mathdetails}

The negative binomial (NB) distribution can be represented as a Poisson--Gamma mixture, where the observed count \( n \) follows a Poisson distribution conditional on a latent rate \( \lambda \), and \( \lambda \) follows a gamma distribution

\begin{equation}
\begin{aligned}
N \sim \text{Pois}(\lambda), \qquad \lambda \sim g\left(r, \frac{\mu}{r}\right),
\end{aligned}
\label{eq:A1}
\end{equation} where 
\(r\) is the dispersion parameter, 
\(\mu\) is the mean parameter, and 
\(\lambda\) is a latent random variable. 
Using the law of total expectation and the law of total variance, the mean and variance of the NB distribution are:
\begin{align*}
E(N) &= E(E(N | \lambda)) = E(\lambda) = \mu,\\
Var(N) &= Var(E(N | \lambda)) + E(Var(N | \lambda))\\
&= Var(\lambda) + E(\lambda)\\
&= \mu + \mu^2 / r.
\end{align*}
This shows that the NB distribution has a mean of $\mu$ and a variance of $\mu + \mu^2 / r$.
The marginal distribution of \( N \) is obtained by integrating over \( \lambda \)
\begin{align*}
\Pr(N = n \mid r, \mu) &= \int_0^\infty \Pr(n \mid \lambda) \Pr(\lambda \mid r, \mu) \, d\lambda\\
&=\int_{0}^{\infty} \frac{e^{-\lambda} \lambda^n}{n!} \frac{(r / \mu)^r}{\Gamma(r)} \lambda^{r-1} \exp\left(-\frac{r \lambda}{\mu}\right) \, d\lambda\\
&= \frac{\Gamma(n + r)}{n! \, \Gamma(r)} \left(\frac{r}{r + \mu}\right)^r \left(\frac{\mu}{r + \mu}\right)^n,
\end{align*}
and turns out to be a negative binomial (NB) model  evaluated at $n$  with dispersion parameter \( r \) and mean \( \mu \).

\section{Derivation of the posterior distribution of AE-level RRs}\label{app:posterior}

The parameters \((p_1, \dots, p_I, \mu_1, \dots, \mu_I, r)\) are obtained by maximizing the likelihood of model~\eqref{ZINB-model}. The AE--level relative reporting rates \(\lambda_{ij}\) are then estimated by empirical Bayes. The prior densities for \(\lambda_{ij}\), 
\begin{equation}
\pi(\lambda_{ij} \mid \hat p_i, \hat \mu_i, \hat r) = \hat p_i \delta_{\lambda_{ij}} + (1 - \hat p_i) g(\lambda_{ij} \mid \hat r, \hat \mu_i / \hat r),
\label{app:prior}
\end{equation}
where the parameters are replaced by their maximum likelihood estimates, are combined with the Poisson likelihood 
\[
L(\lambda_{ij} \mid n_{ij} ) = \frac{\exp(-E_{ij} \lambda_{ij})(E_{ij} \lambda_{ij})^{n_{ij}}}{n_{ij}!}, \quad \lambda_{ij} \geq 0
\]
for $\lambda_{ij}$ given $n_{ij}$.

For $n_{ij} = 0$, the posterior density 
\begin{equation}
\pi(\lambda_{ij} \mid n_{ij} = 0) = \hat{\pi}_{ij} \delta_{\lambda_{ij}} + (1 - \hat{\pi}_{ij}) g\left(\lambda_{ij} \bigm| \hat r, \frac{\hat\mu_i}{\hat r + E_{ij} \hat\mu_i}\right)
\nonumber
\end{equation}
is a mixture of a point mass at $\lambda_{ij}$ and a gamma density, where

\[
\hat{\pi}_{ij} = \Pr(\lambda_{ij} = 0 \mid n_{ij} = 0)
\] represents the posterior probability that $n_{ij} = 0$.
This posterior probability \(\hat{\pi}_{ij}\) is derived via Bayes’ theorem as
\[
\hat{\pi}_{ij} = \Pr(\lambda_{ij} = 0 \mid n_{ij} = 0)=\frac{\Pr(N_{ij} = 0 \mid \lambda_{ij} = 0) \Pr(\lambda_{ij} = 0)}{\Pr(N_{ij} = 0)}.
\]
From the zero--inflated prior, \(\Pr(\lambda_{ij} = 0) = \hat{p}_i\) and, since if \(\lambda_{ij} = 0\), no events occur, \(\Pr(N_{ij} = 0 \mid \lambda_{ij} = 0) = 1\). The marginal probability \(\Pr(n_{ij} = 0)\) integrates over the prior and is given by
\[
\Pr(N_{ij} = 0) = \hat{p}_i + (1 - \hat{p}_i) \int_0^\infty e^{-E_{ij} \lambda_{ij}} g(\lambda_{ij} \mid \hat{r}, \hat{\mu}_i / \hat{r}) d\lambda_{ij},
\]
where the integral is the moment-generating function of the gamma distribution evaluated at \(-E_{ij}\), yielding
\[
\left(\frac{\hat{r}}{\hat{r} + E_{ij} \hat{\mu}_i}\right)^{\hat{r}}.
\]
Combining these components, the posterior probability is
\[
\hat{\pi}_{ij} = \frac{\hat{p}_i}{\hat{p}_i + (1 - \hat{p}_i) \left(\frac{\hat{r}}{\hat{r} + E_{ij} \hat{\mu}_i}\right)^{\hat{r}}}.
\] %
For \(n_{ij} > 0\), the count \(n_{ij} > 0\) excludes the possibility that \(\lambda_{ij} = 0\). Since \(\lambda_{ij} = 0\), then \(\Pr(N_{ij} > 0 \mid \lambda_{ij} = 0) = 0\). Therefore, the posterior probability that \(\lambda_{ij} = 0\) given \(n_{ij} > 0\) is zero and the posterior distribution has no point mass at zero, being fully supported on positive values. Applying Bayes' rule, the posterior distribution is proportional to the product of the Poisson likelihood and the gamma prior,

\[
\pi(\lambda_{ij} \mid n_{ij} > 0) \propto \frac{(E_{ij} \lambda_{ij})^{n_{ij}} e^{-E_{ij} \lambda_{ij}}}{n_{ij}!} \times \frac{\left(\frac{\hat r}{\hat \mu_i}\right)^{\hat r}}{\Gamma(\hat r)} \lambda_{ij}^{\hat r - 1} e^{-\frac{\hat r}{\hat \mu_i} \lambda_{ij}}.
\]

Ignoring constants independent of \(\lambda_{ij}\), this simplifies to

\[
\pi(\lambda_{ij} \mid n_{ij} > 0) \propto \lambda_{ij}^{\hat r + n_{ij} - 1} e^{-\left(E_{ij} + \frac{\hat r}{\hat \mu_i}\right) \lambda_{ij}}.
\]
This is the kernel of a gamma distribution, so the posterior density is

\[
\pi(\lambda_{ij} \mid n_{ij} > 0) = g\left(\lambda_{ij} \bigm| \hat r + n_{ij}, \frac{\hat \mu_i}{\hat r + E_{ij} \hat \mu_i}\right),
\]
which is a gamma distribution with shape parameter \(\hat r + n_{ij}\) and scale parameter \(\hat \mu_i/(\hat r + E_{ij} \hat \mu_i)\).

\section{Derivation of the posterior mean} \label{app:posteriormean}

For \(n_{ij} = 0\), the posterior mean is the expectation of the mixture
\[
\mathbb{E}[\lambda_{ij} \mid n_{ij} = 0] = \hat{\pi}_{ij} \cdot 0 + (1 - \hat{\pi}_{ij}) \times \mathbb{E}\left[\lambda_{ij} \mid \hat{r}, \frac{\hat{\mu}_i}{\hat{r} + E_{ij} \hat{\mu}_i}\right].
\] Since the mean of a gamma distribution \(g(\alpha, \theta)\) is \(\alpha \theta\), this becomes
\[
\mathbb{E}[\lambda_{ij} \mid n_{ij} = 0] = (1 - \hat{\pi}_{ij}) \times \hat{r} \times \frac{\hat{\mu}_i}{\hat{r} + E_{ij} \hat{\mu}_i} = (1 - \hat{\pi}_{ij}) \frac{\hat{r} \hat{\mu}_i}{\hat{r} + E_{ij} \hat{\mu}_i}.
\]
For \(n_{ij} > 0\), the posterior mean is 
\[
\mathbb{E}[\lambda_{ij} \mid n_{ij} > 0] = (\hat{r} + n_{ij}) \frac{\hat{\mu}_i}{\hat{r} + E_{ij} \hat{\mu}_i}.
\]
The empirical Bayes estimate of the AE--level relative reporting rate is the posterior mean of $\lambda_{ij}$,
\begin{equation}
\hat\lambda_{ij} = E(\lambda_{ij} \mid n_{ij}) = 
\begin{cases}
(1 - \hat{\pi}_{ij}) \frac{\hat\mu_i \hat r}{\hat r + E_{ij} \hat\mu_i}, & \text{if } n_{ij} = 0, \\
\frac{\mu_i (\hat r + n_{ij})}{\hat r + E_{ij} \hat\mu_i}, & \text{if } n_{ij} > 0.
\end{cases} 
\end{equation}

\end{document}